\documentclass[10pt,aps,showpacs,nofootinbib,prd,aps,floats,
   222            amsmath,amssymb,amsfonts,floatfix,twocolumn]{revtex4-1}
\usepackage{amsmath, amssymb}
\usepackage{multirow}
\usepackage{paralist}
\usepackage{slashed}
\usepackage{hyperref} % PDF links
\newcommand{\mathsym}[1]{{}}

\usepackage{graphicx}% Include figure files
\usepackage{amsmath}
\usepackage{amssymb}
\usepackage{mathrsfs}
\usepackage{extarrows}
\usepackage{MnSymbol}
\usepackage{graphicx}

\newsavebox{\PSLASH}
 \sbox{\PSLASH}{$p$\hspace{-1.8mm}/}
 
\renewcommand{\theequation}{\thesection.\arabic{equation}}
\newcounter{saveeqn}
\newcommand{\add}{\addtocounter{equation}{1}}
\newcommand{\alphaeqn}{\setcounter{saveeqn}{\value{equation}}%
\setcounter{equation}{0}%
\renewcommand{\theequation}{\mbox{\thesection.\arabic{saveeqn}{\alpha{equation}}}}}
\newcommand{\reseteqn}{\setcounter{equation}{\value{saveeqn}}%
\renewcommand{\theequation}{\thesection.\arabic{equation}}}

 \newsavebox{\notrightarrow}
 \sbox{\notrightarrow}{$\to$\hspace{-4mm}/}
 
 \newsavebox{\PARTIALSLASH}
 \sbox{\PARTIALSLASH}{$\partial$\hspace{-1.6mm}/}
 
 \newsavebox{\ASLASH}
 \sbox{\ASLASH}{$A$\hspace{-2.1mm}/}
 
 \newsavebox{\KSLASH}
 \sbox{\KSLASH}{$k$\hspace{-1.8mm}/}
 
 \newsavebox{\LSLASH}
 \sbox{\LSLASH}{$\ell$\hspace{-1.8mm}/}
 
 \newsavebox{\QSLASH}
 \sbox{\QSLASH}{$q$\hspace{-1.8mm}/}
 
 \newsavebox{\DSLASH}
 \sbox{\DSLASH}{$D$\hspace{-2.2mm}/}
 
 \newsavebox{\DbfSLASH}
 \sbox{\DbfSLASH}{${\mathbf D}$\hspace{-2.8mm}/}
 
 \newsavebox{\DELVECRIGHT}
 \sbox{\DELVECRIGHT}{$\stackrel{\rightarrow}{\partial}$}
 
 \newcommand{\blue}{\IfColor{\textCadetBlue}{}}
\newcommand{\black}{\IfColor{\textBlack}{}}
\newcommand{\red}{\IfColor{\textRed}{}}
\newcommand{\green}{\IfColor{\textOliveGreen}{}}
\newcommand{\lil}{\IfColor{\textRedViolet}{}}

\usepackage{orcidlink}

\begin{document}
\title{Bose-Einstein condensation in a rigidly rotating relativistic boson gas}
\author{E. Siri\,~\orcidlink{0009-0008-1021-3348}~~}\email{e.siri@physics.sharif.ir}\author{N. Sadooghi\,~\orcidlink{0000-0001-5031-9675}~~}\email{Corresponding author: sadooghi@physics.sharif.ir}
\affiliation{Department of Physics, Sharif University of Technology,
P.O. Box 11155-9161, Tehran, Iran}
%%%%%%%%%%%%%%%%%%%%%%%%%%%%%%%%%%%%%%
\begin{abstract}
We study the Bose-Einstein condensation (BEC) of a free Bose gas under rigid rotation. The aim is to explore the impact of rotation on the thermodynamic quantities associated with BEC, including the Bose-Einstein (BE) transition temperature and condensate fraction. We begin by introducing the rotation in the Lagrangian density of free charged Klein-Gordon fields and determine the corresponding grand canonical partition function at finite temperature, chemical potential, and finite angular velocity. Assuming slow rotation, we derive analytical expressions for the pressure, energy, number, and angular momentum densities of a free Bose gas in nonrelativistic and ultrarelativistic limits in terms of the corresponding fugacities. We then focus on the phenomenon of BEC. We calculate the critical temperature of BEC transition and the condensate fraction in a slowly rotating Bose gas including only particles. Our findings indicate that the critical exponent associated with the BE transition in a rotating gas is lower compared to that in a nonrotating Bose gas. We also determine the fugacity in a rotating Bose gas in the aforementioned limits and examine how rotation affects its temperature dependence, both below and above the critical temperature. By analyzing the behavior of heat capacity at these temperatures, we demonstrate that in a nonrelativistic Bose gas, the rotation transforms the nature of the BE phase transition from a continuous to a discontinuous transition.  In general, we find that a nonrelativistic Bose gas under rotation behaves similarly to a nonrotating Bose gas in ultrarelativistic limit.
% subject, objectives, methods, findings, results
\end{abstract}
%\pacs{12.38.Mh, 25.75.-q, 47.65.-d, 52.27.Ny, 5230.Cv }
\maketitle
%-----------------------------------------
\section{Introduction}\label{sec1}
%-----------------------------------------
Studying the properties of matter under extreme conditions is one of the most intriguing areas of modern theoretical and experimental physics \cite{rajagopal2018,pisarski2022}.
It has applications across various branches of physics, ranging from condensed matter to astroparticle physics. These extreme conditions, such as high temperatures and densities, as well as very strong magnetic fields and angular frequencies, are particularly realized in compact stars. These stars are the remnants of massive stars that have experienced gravitational collapse. Among the densest compact stars are neutron stars and white dwarfs, possessing extraordinarily high mass-to-radius ratio. Neutron stars, in particular, are characterized by high densities up to $10^{14} \text{g/cm}^{3}$ in their interior.
This characteristic, combined with the physics of strong interaction, which involves a rich spectrum of particles and attractive interaction channels, allows for pairing of neutrons, protons, mesons, and quarks (see \cite{Pethick:2015jma} for a description of various pairing mechanisms). Among these particles neutrons, protons, and quarks are fermions and mesons are bosons. Regardless of the statistical nature of the pair constituents, these pairings lead to bosonic bound states, that under appropriate circumstances exhibit a BEC. Other research areas related to BEC include weakly interacting dilute gases in magneto-optical traps and strongly interacting quantum gases in optical lattices \cite{Balaz:2010cn}.
\par
As established in statistical mechanics \cite{pathria-book,rebhan-book}, BEC occurs in a nonrelativistic gas of bosons that have been cooled to temperatures very close to absolute zero \cite{Pethick-book,Pitaevski-book}. Consequently, a large fraction of bosons occupy the lowest quantum state. As temperature increases this condensate fraction decreases until it eventually vanishes at a certain critical temperature. The latter is given in terms of the total number density of bosons and their thermal de Broglie wavelength (or thermal length).
This critical temperature corresponds to the point at which the thermal length and the average interparticle distance coincide. At this temperature, the wave functions of individual particles overlap and begin to synchronize. Unexpectedly, this same critical temperature appears to apply to typical temperatures and densities in astrophysical objects, particularly in some neutron and boson stars \cite{gruber}. These stars are self-gravitating compact objects thought to be made of complex scalar or massive vector fields \cite{schunck2003}.
Gravitationally bounded BE condensates are also considered as potential candidates for dark matter \cite{
Sin:1992bg,Boehmer:2007um} (see also \cite{Sharma:2022jio,Aswathi:2023zzn,takeuchi2015,takeuchi2021} and references therein). Recently, the theory of dark matter superfluidity, involving a certain BE transition, has also attracted much attention \cite{Khoury:2021tvy,Lombardo:2000ec}. Both dark matter and meson BE condensates can significantly influence the properties of compact stars. Specifically, the presence of BEC affects their equation of state (EoS) and, consequently, their mass-to-radius relationship. The superfluidity associated with BEC reduces the viscosity of the superfluid component of the star and influences the heat and angular momentum transport within them (see e.g. \cite{manuel2014}). But, apart from extremely large densities, neutron stars are also characterized by their large angular frequencies, which can reach up to $10^3$ Hz. It is known that rotation also affects the thermodynamic properties of neutron stars \cite{hartle1968,sadooghi2023}.
Motivated by the above facts, in this paper, we study the effect of rotation on BEC and the related thermodynamic quantities.
\par
In recent years, extensive research has focused on the effects of rotation on the thermodynamic properties of relativistic systems  \cite{rotation,fukushima2015,rot3,rot5,rot6,rot7,rot8,rot9,rot11,sadooghi2021,Chernodub:2020qah,Fukushima:2024tkz}.
To facilitate the field-theoretical investigation of rotation, it is assumed that the system experiences rigid rotation. Inspired by the notion that large angular velocities, up to $10^{22}$ Hz, are realized in heavy-ion collisions and assuming that the quark-gluon plasma produced in these collisions undergoes rigid rotation \cite{becattini2016}, researchers investigate how rotation affects the thermodynamic properties of fermionic systems. Additionally, the impact of rotation on chiral \cite{sadooghi2021} and confinement/deconfinement phase transitions  \cite{Chernodub:2020qah} is currently a subject of research.
\par
Apart from fermions, the thermodynamic properties of bosons are also affected by rotation. In\cite{chernodub2023-2}, the properties of a spin-one gluon gas are studied under rigid rotation, and the notions of supervorticity and negative Barnett effect are introduced. They are mainly related to the temperature dependence of the moment of inertia of the gluon gas.
The latter is shown to change its sign with increasing temperature. While it is negative below a certain supervortical temperature $T_{s}$, it becomes positive at temperatures larger than $T_{s}$. In \cite{ambrus2023}, it is shown that imaginary angular velocity leads to the appearance of fractal thermodynamics and ninionic statistics in a two dimensional Bose gas bounded to a ring and results in stable ghostlike excitations.
In \cite{sadooghi2024}, the thermodynamic properties of a rigidly rotating Bose gas
are studied. Using the free propagator of a scalar charged field, the full thermodynamic potential, including the zeroth and the first perturbative and nonperturbative
ring corrections, is determined. The full pressure arising from this calculation yields various thermodynamic quantities associated with a rigidly rotating Bose gas. In particular, it is shown that under certain circumstance, negative moment of inertia and superluminal sound velocities arise. These properties are interpreted as thermodynamic instabilities appearing at high temperatures in a strongly interacting Bose gas under rigid rotation.
\par
In the present paper, we study the effect of rigid rotation on BEC in an ideal Bose gas. We determine first the thermal pressure and the EoS of a slowly rotating Bose gas in two nonrelativistic (NR) and ultrarelativistic (UR) limits. In both cases, we focus solely on the contribution of particles, allowing for a direct comparison of the corresponding results in these two limits. This assumption can only be made in systems in which a charge imbalance exists. We use the same field theoretical methods as are introduced in \cite{kapusta1979,haber1981,haber1982, kapusta-book,shabad1987}.
We show that in both limits, rigid rotation decreases the speed of sound in the corotating frame of the rotating gas. We then focus on BEC and determine, in particular, its critical temperature and the condensate fraction. We demonstrate that comparing to the critical temperature of BE transition of a nonrotating Bose gas, the critical temperature of a rotating gas is lower, although by increasing the angular velocity the transition temperature increases. We determine the dependence of the fugacities associated with a rotating Bose gas in NR and UR limits on temperature and angular velocities. Using these results, we derive analytical expressions for the pressure, heat capacity, and angular momentum density at temperatures both below and above the critical temperature of BEC. According to our results, the thermodynamic properties of a rotating Bose gas in nonrelativistic limit is similar to an ultrarelativistic gas in the absence of rotation. In particular, the temperature dependence of the heat capacity in a nonrelativistic Bose gas under rotation exhibits a discontinuity at the critical temperature, similar to the one that occurs in ultrarelativistic Bose gases without rotation.
We compute the latent heat necessary to convert the condensate to thermal phase at the critical temperature and show that in a rotating Bose gas, it is, in general, larger than the latent heat of a Bose gas in the absence of rotation. It is important to note that the effect of fast rotation on BEC that occurs in an anharmonic trap has been previously studied in \cite{cond2006}. The anharmonicity is introduced by a potential including a quadratic and a quartic terms. Although the main framework and the way the rotation is introduced in \cite{cond2006} is completely different from the method used in the present paper, our findings partly coincide with the results presented in \cite{cond2006}.
\par
The organization of this paper is as follows: In Sec. \ref{sec2}, we determine the free propagator of a rotating ideal Bose gas at temperature $T$, chemical potential $\mu$, and angular velocity $\Omega$. Using field theoretical methods, we then compute the corresponding partition function and pressure of a rotating ideal Bose gas. In Sec. \ref{sec3}, the thermal pressure and the EoS of a slowly rotating Bose gas is determined in NR and UR limits.  In Sec. \ref{sec4}, we focus on the BE transition, determine its critical temperature, and how it is affected by small rotation. In Sec. \ref{sec5}, we determine the $T$ dependence of the fugacity associated to BEC at temperatures lower and higher than the critical temperature. Using these results, we determine the pressure, heat capacity, and angular momentum density of a rotating Bose gas in NR and UR limits.
Section \ref{sec6} is devoted to concluding remarks. In Appendices \ref{appA} and \ref{appB} a number of useful formula and a method for solving certain integrals are introduced.
%-----------------------------------------
\section{The pressure of a rotating ideal Bose gas}\label{sec2}
\setcounter{equation}{0}
%-----------------------------------------
We start with the Lagrangian density of free complex scalar fields $\phi$,
	\begin{eqnarray}\label{A1}
	\mathcal{L}	=g^{\mu\nu} \partial_{\mu}\phi^{\star} \partial_{\nu}\phi-m^{2} \phi^{\star}\phi,
	\end{eqnarray}
and introduce the rigid rotation by the metric
	\begin{eqnarray}\label{A2}
			{{g}_{\mu \nu }}=\left( \begin{matrix}
			1-(x^2+y^2)\Omega^2 & y\Omega  & -x\Omega  & 0  \\
			y\Omega  & -1 &0 & 0  \\
			-x\Omega  & 0& -1 & 0  \\
			0 & 0 & 0 & -1  \\
		\end{matrix} \right).
	\end{eqnarray}
Here, $m$ is the rest mass and $x$ as well as $y$ are components of space-time coordinates $x^{\mu}=(t,x,y,z)$ and $\Omega$ is the constant angular velocity for a rotation around the $z$ axis.
Plugging the above metric into \eqref{A1}, the resulting Lagrangian density reads
\begin{eqnarray}\label{A3}
\mathcal{L}= (\partial_{0}-i\Omega L_{z}) \phi^{\star}(\partial_{0}-i\Omega L_{z}) \phi - \left|\boldsymbol{\nabla}\phi \right|^{2} - m^2 \left| \phi \right|^{2},\nonumber\\
\end{eqnarray}
where $L_{z}$ is the $z$-component of the angular momentum, defined by $L_{z}\equiv i(y\partial_{x}-x\partial_{y})$. The above Lagrangian is invariant under global $U(1)$ transformation,
\begin{eqnarray}\label{A4}
\phi(x)&\to&e^{-i\alpha}\phi(x),\nonumber\\
\phi^{\star}(x)&\to&e^{ i\alpha}\phi^{\star}(x).
\end{eqnarray}
The corresponding conserved charge density is given by
\begin{eqnarray}\label{A5}
j^{0}=-i\left(\pi^{\star}\phi-\pi\phi^{\star}\right),
\end{eqnarray}
where the conjugate momenta read
		\begin{eqnarray}\label{A6}
			\pi&=&\partial_{0}\phi-i\Omega L_{z} \phi ,\nonumber\\
			\pi^{\star} &=&\partial_{0}\phi^{\star}-i\Omega L_{z} \phi^{\star}.
		\end{eqnarray}
At finite temperature $T$ and chemical potential $\mu$, the partition function $\mathcal{Z}$ of this model is obtained from
\begin{eqnarray}\label{A7}
\mathcal{Z} = \int \mathcal{D}\pi^{\star}\mathcal{D}\pi \int \mathcal{D}\phi^{\star}\mathcal{D}\phi \exp \left(\int_{X} \mathcal{J}[\pi,\pi^{\star};\phi,\phi^{\star}] \right),\nonumber\\
\end{eqnarray}
where
\begin{eqnarray}\label{A8}
\int_{X} \equiv \int_{0}^{\beta} d\tau \int d^{3}x,
\end{eqnarray}
with the imaginary time $\tau=it$, $\beta\equiv T^{-1}$, and
\begin{eqnarray}\label{A9}
\mathcal{J}[\pi,\pi^{\star};\phi,\phi^{\star}]\equiv \pi^{\star}\partial_{0}\phi+\pi\partial_{0}\phi^{\star}-\mathcal{H}+\mu j^{0}.
\end{eqnarray}
Here, $j^{0},\pi,$ and $\pi^{\star}$ are defined in \eqref{A5} and \eqref{A6}, and the Hamiltonian density $\mathcal{H}$ is given by
\begin{eqnarray}\label{A10}
\mathcal{H}=\partial_{0}\phi^{\star}\pi+\partial_{0}\phi \pi^{\star} - \mathcal{L}.
\end{eqnarray}
Plugging $\mathcal{L}$ into \eqref{A10} and using \eqref{A6}, $\mathcal{H}$ can be rewritten as
\begin{eqnarray}\label{A11}
\mathcal{H}= \pi^{\star}\pi+i\Omega(\pi L_{z} \phi^{\star}+\pi^{\star} L_{z} \phi) + \left| \boldsymbol{\nabla}{\phi} \right|^{2} + m^{2} \left| \phi \right|^{2}.\nonumber\\
\end{eqnarray}
To integrate over $\pi$ and $\pi^{\star}$ in \eqref{A7}, we introduce the shifted momenta $\tilde{\pi}$ and $\tilde{\pi}^{\star}$ as
\begin{eqnarray}\label{A12}
\tilde{\pi} &=& \pi-\partial_{0}\phi+i\Omega L_{z}\phi+i \mu \phi,\nonumber \\	
\tilde{\pi}^{\star} &=& \pi^{\star}-\partial_{0}\phi^{\star}+i\Omega L_{z}\phi^{\star} -i \mu \phi^{\star},
\end{eqnarray}
use \eqref{A1} and \eqref{A11}, and arrive eventually at
\begin{eqnarray}\label{A13}
\mathcal{J}[\pi,\pi^{\star};\phi,\phi^{\star}]=\mathcal{L}^{\prime}-\tilde{\pi}^{\star}\tilde{\pi},
\end{eqnarray}
where $\mathcal{L}^{\prime}$ is given by
\begin{eqnarray}\label{A14}
\hspace{-0.5cm}\mathcal{L'} = \left| (\partial_{0}-i\Omega L_{z}-i\mu) \phi \right|^{2} - \left| \boldsymbol{\nabla}{\phi} \right|^{2} - m^{2} \left| \phi \right|^{2}.
\end{eqnarray}
Plugging \eqref{A13} into \eqref{A7} and performing the Gaussian integration over $\tilde{\pi}$ and $\tilde{\pi}^{\star}$, the partition function reads
\begin{eqnarray}\label{A15}
\mathcal{Z} = {N} \int \mathcal{D}\phi^{\star}\mathcal{D}\phi \exp \left[\int_{X} \mathcal{L}^{\prime} \right],
\end{eqnarray}
with a factor ${N}$ arising from the integration over $\tilde{\pi}$ and $\tilde{\pi}^{\star}$, and $\mathcal{L}^{\prime}$ from \eqref{A14}. To perform the integration over $\phi$ and $\phi^{\star}$, we introduce the mode expansion \cite{siri2024-1}
\begin{eqnarray}\label{A16}
\phi(x)=\zeta+\sqrt{\beta V}\sumint_{n,\ell \neq 0,k} e^{i(\omega_{n}\tau+\ell \varphi + k_{z}z)} J_{\ell}(k_{\perp}r) \tilde{\phi}_{n,\ell}(k),\nonumber\\
\end{eqnarray}
where we have separated the zero-mode contribution $\zeta$ in order to study the BEC phenomenon in a rigidly rotating Bose gas \cite{kapusta-book}. Moreover, in order to implement a $U(1)$ symmetry breaking, we choose $\zeta$ to be real. Because of the cylindrical symmetry, we use in \eqref{A16} a cylindrical coordinate system described by $x^{\mu}=(t,x,y,z)=(t,r\cos\varphi,r\sin\varphi,z)$, with $r$ the radial coordinate, $\varphi$ the azimuthal angle, and $z$ the height of the cylinder. The corresponding conjugate momenta at finite temperature are given by Matsubara frequency $\omega_{n}=2\pi nT$, discrete quantum number $\ell$, the eigenvalue of $L_{z}$, and continuous momenta $k_{z}$ and $k_{\perp}\equiv (k_{x}^{2}+k_{y}^{2})^{1/2}$ in cylindrical coordinates. The latter appears in the Bessel function $J_{\ell}(k_{\perp} r)$. In \eqref{A16}, an appropriate Fourier-Bessel transformation is used with
\begin{eqnarray}\label{A17}
\sumint_{n,\ell \neq 0,k}\equiv\sum_{n\neq 0}\sum_{\ell\neq 0}\int d\tilde{k},\qquad \text{and}\qquad d\tilde{k}\equiv \frac{dk_{\perp}k_{\perp}dk_{z}}{(2\pi)^{2}}. \nonumber\\
\end{eqnarray}
Plugging, at this stage, the mode expansion of $\phi$ from \eqref{A16} into $\mathcal{L}^{\prime}$ in \eqref{A15}, and performing the ingeration over the cylindrical coordinates in
\begin{eqnarray}\label{A18}
\int_{X}=\int_{0}^{\beta}d\tau\int_{0}^{\infty}r dr\int_{0}^{2\pi}d\varphi\int_{-\infty}^{ \infty}dz,
\end{eqnarray}
by making use of \eqref{appA2} and \eqref{appA4} as well as \eqref{appA5}, we arrive at
\begin{eqnarray}\label{A19}
\lefteqn{\hspace{-0.5cm}\int_{X} \mathcal{L'}=\frac{V}{T} (\mu^{2}-m^{2})\zeta^{2}}\nonumber\\
&&- V \sumint_{n,\ell \ne 0,k} \tilde{\phi}^{\star}_{n,\ell}(k) \, [\beta^{2} D_{\ell,0}^{-1}(k)] \, \tilde{\phi}_{n,\ell}(k),
\end{eqnarray}
where the inverse free propagator $D_{\ell,0}^{-1}(k)$ reads
\begin{eqnarray}\label{A20}
D_{\ell,0}^{-1}(k) \equiv (\omega_{n}+i\mu_{\text{eff}})^{2}+\omega^{2}.
\end{eqnarray}
In \eqref{A20}, $\mu_{\text{eff}}\equiv \mu+\ell\Omega$ is the effective chemical potential,
 including the chemical potential and the angular velocity, and $\omega^{2}\equiv k_{\perp}^{2}+k_{z}^{2}+m^{2}$. Let us notice that in \cite{sadooghi2024}, the free propagator of a rotating Bose gas at finite temperature and zero chemical potential is determined using the Fock-Schwinger method. Setting $\mu=0$ in \eqref{A20}, we arrive at the same result as in \cite{sadooghi2024}. Plugging, at this stage, \eqref{A19} into \eqref{A15} and performing the functional integral over $\phi$ and $\phi^{\star}$, we arrive first at
\begin{eqnarray}\label{A21}
\ln\mathcal{Z}=\beta V\left(\mu^{2}-m^{2}\right)\zeta^{2}-\frac{1}{2}\ln\left|\beta^{2}D_{\ell,0}^{-1}(k)\right|.
\end{eqnarray}
Following standard steps as described e.g. in \cite{kapusta-book}, we then obtain
\begin{eqnarray}\label{A22}
\ln\mathcal{Z}&=&\beta V\left(\mu^{2}-m^{2}\right)\zeta^{2}-\frac{V}{2}\sum\limits_{n=-\infty}^{ \infty}\sum\limits_{\ell\neq 0}\sum_{\epsilon=\pm}\int d\tilde{k}\nonumber\\
&&\times
\ln \left[\beta^{2}\left( \omega _{n}^{2}+\left( \omega +\epsilon\mu_{\text{eff}}  \right)^{2} \right) \right].
\end{eqnarray}
Performing the sum over Matsubara frequencies by making use of
\begin{eqnarray}\label{A23}
\hspace{-0.5cm}\sum\limits_{n=-\infty }^{ \infty }\ln \left(\left( 2n\pi  \right)^{2}+u^{2} \right)=u+2\ln \left( 1-e^{-u} \right),
\end{eqnarray}
we arrive finally at the partition function of a relativistic free Bose gas under rigid rotation,
\begin{eqnarray}\label{A24}
\ln \mathcal{Z} &=& \beta V(\mu^{2}-m^{2})\zeta^{2} -\beta V \sum\limits_{\ell\neq 0}\int d\tilde{k}~ \big\{\omega \nonumber\\
&&+T\big[\ln\left(1-e^{-\beta (\omega - \mu_{\text{eff}})}\right)+\ln\left(1-e^{-\beta (\omega + \mu_{\text{eff}})}\right)]\big\}.\nonumber\\
\end{eqnarray}
The corresponding pressure $P$ is then given by
\begin{eqnarray}\label{A25}
P&=&\frac{\ln \mathcal{Z}}{\beta V}= (\mu^{2}-m^{2})\zeta^{2}-\sum\limits_{\ell\neq 0}\int d\tilde{k}~\big\{\omega \nonumber\\
&&+T\big[\ln\left(1-e^{-\beta (\omega - \mu_{\text{eff}})}\right)+\ln\left(1-e^{-\beta (\omega + \mu_{\text{eff}})}\right)\big]\big\}.\nonumber\\
\end{eqnarray}
The first term on the right hand side (rhs) of \eqref{A25} is the contribution from the zero mode characterized by $\boldsymbol{k}=\boldsymbol{0}$ and $\ell=0$, the second term, including $\omega$ is the zero temperature part and last two terms correspond to the thermal part of the pressure. Using the thermal part, it is possible to determine other thermodynamic quantities associated with a rotating Bose gas.
Using the Gibbs-Duhem relation $dP=sdT+nd\mu+jd\Omega$ \cite{chernodub2023-2}, we arrive at the entropy density $s$, the number (charge) density $n$, angular momentum density $j$,
\begin{eqnarray}\label{A26}
s=\left(\frac{\partial P}{\partial T}\right)_{\mu,\Omega},~
n=\left(\frac{\partial P}{\partial \mu}\right)_{T,\Omega},~
j=\left(\frac{\partial P}{\partial \Omega}\right)_{T,\mu}.
\end{eqnarray}
In a rotating gas, we have to distinguish between the energy density of the corotating frame $\epsilon$ and the energy density in the nonrotating laboratory frame $\epsilon^{\text{lab}}$ \cite{landau-book}. Their relation is given by $\epsilon=\epsilon^{\text{lab}}-j\Omega$. In this work, we determine $\epsilon$ from
\begin{eqnarray}\label{A27}
\epsilon=T^{2}\frac{\partial}{\partial T}\left(\frac{P}{T}\right)_{z},
\end{eqnarray}
where $z$ is the fugacity. Adding the Gibbs-Duhem relation $dP=sdT+nd\mu+jd\Omega$ with $d\epsilon=Tds+\mu dn-jd\Omega$, arising also in \cite{chernodub2023-2}, we arrive immediately at $\epsilon+p=Ts+\mu n$. The entropy density is thus given by
\begin{eqnarray}\label{A28}
s=\beta\left(\epsilon+P-n\mu\right).
\end{eqnarray}
Let us notice that \eqref{A27} leads to the EoS $\epsilon=\epsilon(P)$ of the Bose gas in the corotating frame, which eventually yields the speed of sound $c_{s}^{2}\equiv \frac{\partial P}{\partial \epsilon}$ in this medium. Finally, the heat capacity $C_{V}$ is obtained by utilizing
\begin{eqnarray}\label{A29}
C_{V}\equiv\left(\frac{\partial\epsilon}{\partial T}\right)_{n,\Omega}.
\end{eqnarray}
Setting $\Omega=0$ in the above relations, we arrive at the corresponding expressions to  thermodynamic quantities of a nonrotating Bose gas.
\par
In Sec. \ref{sec3}, we determine the general expression for the pressure of a rotating Bose gas in NR and UR limits. For simplicity, we introduce the notation\footnote{In following sections, we neglect the zero temperature contribution to $P$.}
\begin{eqnarray}\label{A30}
P=P_{0}+P_{\text{th}},
\end{eqnarray}
where
\begin{eqnarray}\label{A31}
P_{0}\equiv\left(\mu^{2}-m^{2}\right)\zeta^{2},
\end{eqnarray}
is the zero mode contribution to $P$, and $P_{\text{th}}\equiv P^{+}+P^{-}$ consisting of
\begin{eqnarray}\label{A32}
P^{\pm}\equiv -T\sum_{\ell\neq 0}\int d\tilde{k}\ln\left(1-e^{\pm \beta\mu}e^{-\beta(\omega\mp\ell\Omega)}\right),
\end{eqnarray}
the thermal part of $P$. The superscript $\pm$ denotes the thermal contributions of particles ($+$) and antiparticles ($-$) to $P$. In this paper, we focus only on $P^{+}$ and determine it in NR and UR limits. Using \eqref{A27}, it is also possible to compute the energy density $\epsilon^{+}$. We compare our results with the standard thermal pressure and energy density of a nonrotating Bose gas in these two limits. Using these results, we determine in Sec. \ref{sec4}, the number (charge) density $n^{+}$ and discuss the BEC phenomenon for a rotating Bose gas in both cases. Assuming slow rotation ($\Omega\ll T$), it is possible to analytically determine the critical temperature $T_{c}$ of BEC and explore the impact of $\Omega$ on $T_{c}$ in these limits. In Sec. \ref{sec5}, we first turn back to the pressure in NR and UR limits and determine it for $T\leq T_{c}$ and $T>T_{c}$. We then compute the fugacity, heat capacity, angular momentum density, and latent heat at $T=T_c$. In each step, we compare our results with the standard results of a nonrotating Bose gas. Since we only consider the contribution of particles and neglect the contributions of antiparticles, we skip the subscript $"+"$ in $P^{+}$ and other thermodynamic quantities arising from $P$.
%-----------------------------------------
\section{Thermal pressure of a slowly rotating Bose gas}\label{sec3}
\setcounter{equation}{0}
%-----------------------------------------
\subsection{Nonrelativistic limit}\label{sec3a}
%-----------------------------------------
Let us consider the thermal part of the pressure $P$ from \eqref{A32}. In NR limit,
$\omega=\left(\boldsymbol{k}^{2}+m^{2}\right)^{1/2}$ with $\boldsymbol{k}^{2}\equiv k_{z}^{2}+k_{\perp}^{2}$ is approximated by $\omega\sim\omega_{k}+m$, with $\omega_{k}\equiv\frac{k^{2}}{2m}$ and $k=|\boldsymbol{k}|$ \cite{kapusta-book, haber1982}. Introducing $z$ in terms of $T$ and $\Omega$ dependent chemical potential $\mu$,
\begin{eqnarray}\label{B1}
z\equiv e^{\beta(\mu-m)},
\end{eqnarray}
the pressure $P^{ }$ in NR limit is thus given by
\begin{eqnarray}\label{B2}
P_{\text{nr}}\equiv -T\sum_{\ell\neq 0}\int d\tilde{k}\ln\left(1-z e^{-\beta(\omega_{k}-\ell\Omega)}\right).
\end{eqnarray}
The integral converges only for $z e^{-\beta (\omega_{k}-\ell\Omega)}<1$. This condition determines the signs of $\mu$ and $\ell$. On the one hand, $z<1$ leads to $\mu< m$ \cite{kapusta-book}, on the other hand, assuming $\Omega>0$, the condition $e^{-\beta(\omega_{k}-\ell\Omega)}$ leads to $\ell<\text{min}(\omega_{k})=0$. We thus arrive first at
\begin{eqnarray}\label{B3}
P_{\text{nr}}&=&-T\sum_{\ell=-\infty}^{-1}\int d\tilde{k}\ln\left(1-ze^{-\beta (\omega_{k}-\ell\Omega)}\right).
\end{eqnarray}
Performing then a change of variable $\ell\to -\ell$ in $P_{\text{nr}}$, we obtain
\begin{eqnarray}\label{B4}
P_{\text{nr}}= -T\sum_{\ell=1}^{ \infty}\int d\tilde{k}\ln\left(1-ze^{-\beta(\omega_{k}+\ell\Omega)}\right).
\end{eqnarray}
To compute $P_{\text{nr}}$, we have to perform the integration over $k$ and the sum over $\ell$. To evaluate the momentum integration, we use
\begin{eqnarray}\label{B5}
\ln(1-x)=-\sum_{j=1}^{ \infty}\frac{x^{j}}{j}, \quad \text{for}\quad x<1.
\end{eqnarray}
This is possible because $x=ze^{-\beta(\omega_{k}+\ell\Omega)}<1$. We thus arrive at
\begin{eqnarray}\label{B6}
P_{\text{nr}}=T\sum_{j=1}^{ \infty}\frac{z^{j}}{j}\sum_{\ell=1}^{ \infty}e^{-\beta j\ell\Omega} \int d\tilde{k}~e^{-\beta j\omega_{k}}.
\end{eqnarray}
For $\beta j\ell\Omega>0$, the summation over $\ell$ can be carried out and yields
\begin{eqnarray}\label{B7}
\sum_{\ell=1}^{ \infty}e^{-\beta j\ell\Omega}=\frac{1}{1-e^{-\beta j\Omega}}.
\end{eqnarray}
The momentum integration in \eqref{B6} can also be performed by following the method described in Appendix \ref{appB1} and leads to
\begin{eqnarray}\label{B8}
\int d\tilde{k}~e^{-\beta j\omega_{k}}=\frac{1}{\lambda_{T}^{3}j^{3/2}},
\end{eqnarray}
with $\lambda_{T}$ defined by
\begin{eqnarray}\label{B9}
\lambda_{T}\equiv \left(\frac{2\pi}{mT}\right)^{1/2}.
\end{eqnarray}
Plugging \eqref{B7} and \eqref{B8} into \eqref{B6}, we first arrive at
\begin{eqnarray}\label{B10}
P_{\text{nr}}=\frac{T}{\lambda_{T}^{3}}\sum_{j=1}^{ \infty}\frac{z^{j}}{j^{5/2}}\frac{1}{(1-e^{-\beta j\Omega})}.
\end{eqnarray}
In a slowly rotating Bose gas with $\beta\Omega\ll 1$, where according to
\begin{eqnarray}\label{B11}
\frac{1}{1-e^{-x}}\xrightarrow{x\ll 1}\frac{1}{x}\qquad\mbox{for}\qquad x>0,
\end{eqnarray}
the factor $(1-e^{-\beta j\Omega})^{-1}$ in \eqref{B10} behaves as $(\beta j\Omega)^{-1}$,
 $P_{\text{nr}}$ is given by
\begin{eqnarray}\label{B12}
P_{\text{nr}}\simeq \frac{T}{\lambda_{T,\Omega}^{3}}g_{7/2}(z).
\end{eqnarray}
Here, the modified thermal length in a slowly rotating Bose gas is defined by $\lambda_{T,\Omega}\equiv \lambda_{T}(\beta\Omega)^{1/3}$ with $\lambda_{T}$ from \eqref{B9}. Moreover, the BE function $g_{s}(z)$ of order $s$ is given by\footnote{In general, $g_{s}(e^{\alpha})$ are defined for $\alpha\leq 0$ leading to $z=e^{\alpha}\leq 1$. For $\alpha=0$, $g_{s}(1)=\zeta(s)$, where $\zeta(s)$ the Riemann Zeta-function. Note that without the condition $\alpha\leq 0$, the integral appearing in \eqref{B12} is equal to the polylogarithm function $\mbox{Li}_{s}(z)$ defined by  $\mbox{Li}_{s}(z)\equiv \sum_{j=1}^{\infty}z^{j}/j^{s}$.}
\begin{eqnarray}\label{B13}
g_{s}(z)\equiv\frac{1}{\Gamma(s)}\int_{0}^{\infty}\frac{x^{s-1} dx}{ze^{x}-1},
\end{eqnarray}
with $z=e^{\beta(\mu-m)}$ and $\mu<m$.
\par
Plugging at this stage $P_{\text{nr}}$ from \eqref{B12} into \eqref{A26}, it is possible to derive the thermal part of the number (charge) density,
\begin{eqnarray}\label{B14}
n_{\text{nr}}&=&\frac{1}{\lambda_{T,\Omega}^{3}}g_{5/2}(z).
\end{eqnarray}
In the next section, we combine $n_{\text{nr}}$ with the contribution of $P_{0}$ from \eqref{A31} to the total number density $n_{\text{tot}}$ and study the BEC phenomenon in a slowly rotating Bose gas in NR limit. We show that $P_{\text{nr}}$ exhibits different behavior below and above a certain BE transition temperature. But, before discussing this interesting phenomenon, let us substitute $P_{\text{nr}}$ from \eqref{B12} into \eqref{A27} to  arrive at the (internal) energy density $\epsilon$,
\begin{eqnarray}\label{B15}
\epsilon_{\text{nr}}=\frac{5}{2} P_{\text{nr}}.
\end{eqnarray}
This reveals the EoS of a slowly rotating Bose gas in NR limit. According to this result, the sound velocity in this case is $c_{s}=(2/5)^{1/2}\sim 0.63$.
\par
Let us notice that the above results are only valid for $\Omega\neq 0$. Obviously, the pressure of a nonrotating Bose gas in NR limit is given by \cite{pathria-book,rebhan-book}
\begin{eqnarray}\label{B16}
P_{\text{nr}}^{(0)}=-T\int\frac{d^{3}k}{(2\pi)^{3}}\ln\left(1-ze^{-\beta\omega_{k}}\right)\simeq \frac{T}{\lambda_{T}^{3}}g_{5/2}(z).\nonumber\\
\end{eqnarray}
Moreover, the thermal part of the number density reads
 \begin{eqnarray}\label{B17}
n_{\text{nr}}^{ (0)}=\frac{1}{\lambda_{T}^{3}}g_{3/2}\left(z\right).
\end{eqnarray}
Plugging \eqref{B16} into \eqref{A27}, the EoS of an ideal nonrotating Bose gas in NR limit is, as expected, given by
\begin{eqnarray}\label{B18}
\epsilon_{\text{nr}}^{ (0)}=\frac{3}{2} P_{\text{nr}}^{ (0)},
\end{eqnarray}
with $ P_{\text{nr}}^{ (0)}$ from \eqref{B16}. This EoS yields $c_{s}^{(0)}=(2/3)^{1/2}\sim 0.82$, in contrast to the speed of sound $c_{s}\sim 0.63$ in a slowly rotating nonrelativistic Bose gas. It turns out that in this case, the speed of sound is smaller than in a nonrotating gas.
%-----------------------------------------
\subsection{Ultrarelativistic limit}\label{sec3b}
%-----------------------------------------
We consider the thermal part of the pressure $P$ from \eqref{A32}. In UR limit, we replace $\omega=\left(\boldsymbol{k}^{2}+m^{2}\right)^{1/2}$ with $\omega_{k}\equiv k$. Introducing the fugacity
\begin{eqnarray}\label{B19}
\mathfrak{z}\equiv e^{\beta\mu},
\end{eqnarray}
the pressure $P$ in UR limit is given by
\begin{eqnarray}\label{B20}
P_{\text{ur}}\equiv -T\sum_{\ell\neq 0}\int d\tilde{k}\ln\left(1-\mathfrak{z} e^{-\beta(\omega_{k}-\ell\Omega)}\right).
\end{eqnarray}
Note that as in NR limit, the fugacity depends explicitly on $T$ and $\Omega$. In Sec. \ref{sec5}, this dependence will be explicitly determined [see \eqref{D27} and Fig. \ref{fig5}].
Following the same arguments as in Sec. \ref{sec3a} for the NR limit, the integral appearing in \eqref{B20} is convergent only if the chemical potential $\mu<0$ and $\ell<0$. After performing a change of variable $\ell\to -\ell$, the thermal pressure in UR limit is thus given by
\begin{eqnarray}\label{B21}
P_{\text{ur}}= -T\sum_{\ell=1}^{ \infty}\int d\tilde{k}\ln\left(1-\mathfrak{z}e^{-\beta(\omega_{k}+\ell\Omega)}\right).
\end{eqnarray}
Using \eqref{B5}, the pressure $P_{\text{ur}}$ reads
\begin{eqnarray}\label{B22}
P_{\text{ur}}=T\sum_{j=1}^{ \infty}\frac{\mathfrak{z}^{j}}{j}\sum_{\ell=1}^{ \infty}e^{-\beta j\ell\Omega}\int d\tilde{k}~e^{\beta j\omega_{k}}.
\end{eqnarray}
As in the previous case, for $\Omega>0$, the summation over $\ell$ yields \eqref{B7}. To perform the $k$-integration, we follow the arguments in Appendix \ref{appB2} [see \eqref{appB14}] and arrive at
\begin{eqnarray}\label{B23}
\int d\tilde{k}~e^{-\beta j\ell\Omega}=\frac{1}{\Lambda_{T}^{3}j^{3}},
\end{eqnarray}
where $\Lambda_{T}$ is defined by
\begin{eqnarray}\label{B24}
\Lambda_{T}\equiv \frac{\pi^{2/3}}{T}.
\end{eqnarray}
Plugging \eqref{B7} and \eqref{B23} into \eqref{B22}, we obtain
\begin{eqnarray}\label{B25}
P_{\text{ur}}=\frac{T}{\Lambda_{T}^{3}}\sum_{j=1}^{ \infty}\frac{\mathfrak{z}^{j}}{j^{4}}\frac{1}{(1-e^{-\beta j\Omega})}.
\end{eqnarray}
For slow rotation with $\beta\Omega\ll 1$, we use \eqref{B11} and arrive at
\begin{eqnarray}\label{B26}
P_{\text{ur}}^{ }\simeq \frac{T}{\Lambda_{T,\Omega}^{3}}g_{5}(\mathfrak{z}).
\end{eqnarray}
Here, the BE function $g_{s}(\mathfrak{z})$ of order $s=5$ is given in \eqref{B13} and the modified thermal length for an UR Bose gas under rotation is defined by $\Lambda_{T,\Omega}\equiv \Lambda_{T}(\beta\Omega)^{1/3}$ with $\Lambda_{T}$ from \eqref{B24}.
\par
Similar to the previous case, the number (charge) density $n_{\text{ur}}$ is given by
\begin{eqnarray}\label{B27}
n_{\text{ur}}&=&\frac{1}{\Lambda_{T,\Omega}^{3}}g_{4}\left(\mathfrak{z}\right).
\end{eqnarray}
Moreover, the energy density arises from \eqref{A27} with $P$ replaced with $P_{\text{ur}}$ from \eqref{B26},
\begin{eqnarray}\label{B28}
\epsilon_{\text{ur}}=4P_{\text{ur}},
\end{eqnarray}
leading to $c_{s}=0.50$.
\par
The above results in UR limit are different from the ones in a nonrotating ultrarelativistic Bose gas, where the corresponding pressure is given by \cite{pathria-book}
 \begin{eqnarray}\label{B29}
P_{\text{ur}}^{ (0)}=-T\int\frac{d^{3}k}{(2\pi)^{3}}\ln\left(1-\mathfrak{z}e^{-\beta\omega_{k}}\right)\simeq\frac{T}{\Lambda_{T}^{3}}g_{4}\left(\mathfrak{z}\right), \nonumber\\
\end{eqnarray}
with $\mathfrak{z}$ from \eqref{B19} and $\Lambda_{T}$ from \eqref{B24}. In this case, the corresponding number (charge) density reads
\begin{eqnarray}\label{B30}
n_{\text{ur}}^{ (0)}&=&\frac{1}{\Lambda_{T}^{3}}g_{3}\left(\mathfrak{z}\right).
\end{eqnarray}
Plugging \eqref{B29} into \eqref{A27}, the energy density of a nonrotating $\epsilon_{\text{ur}}^{(0)}$ is given by
\begin{eqnarray}\label{B31}
\epsilon^{(0)}_{\text{ur}}=3P^{ (0)}_{\text{ur}}.
\end{eqnarray}
This result is in contrast to \eqref{B28} and leads to the speed of sound $c_{s}^{(0)}=(1/3)^{1/2}\sim 0.58$ in a nonrotating ultrarelativistic Bose gas. Similar to the NR limit, in the UR limit, the speed of sound in a rotating gas is lower than that in a nonrotating Bose gas in the same limit. Moreover, according to the above results, a nonrelativistic Bose gas under rigid rotation behaves as an ultrarelativistic Bose gas in the absence of rotation. Additional evidence supporting this statement is provided in following sections.
%-----------------------------------------
\section{Bose-Einstein condensation}\label{sec4}
\setcounter{equation}{0}
%-----------------------------------------
According to the descriptions of previous sections, the total pressure \eqref{A25} has the following structure
\begin{eqnarray}\label{C1}
P=P_{0}+P_{\text{th}},
\end{eqnarray}
where the zero mode part of the pressure, $P_{0}=\left(\mu^{2}-m^{2}\right)\zeta^{2}$ is defined in \eqref{A31} and $P_{\text{th}}=P$ is originally defined in \eqref{A32}.
In Sec. \ref{sec3}, we determined $P$, consisting only of the contribution of particles, in two NR and UR limits [see \eqref{B12} and \eqref{B26}]. Using the definition of the number density from \eqref{A26}, we also determined the thermal part of $n$, denoted by $n_{\text{th}}$, in these two limits. The total number density is thus given by
\begin{eqnarray}\label{C2}
n_{\text{tot}}=n_{0}+n_{\text{th}},
\end{eqnarray}
where
\begin{eqnarray}\label{C3}
n_{0}=2\mu\zeta^{2},
\end{eqnarray}
and $n_{\text{th}}$ of a rotating (nonrotating) Bose gas is given by $n_{\text{nr}}$ ($n^{ (0)}_{\text{nr}}$) and $n_{\text{ur}}$ ($n^{ (0)}_{\text{ur}}$) in the NR and UR limits.
\par
In this section, we focus on the condensate $\zeta$ and determine it at temperature below and above the BEC critical temperature. Following the arguments from \cite{kapusta-book}, the condition under which $\zeta$ is nonzero is given by minimizing $P$ (or the partition function $\ln\mathcal{Z}$) with respect to $\zeta$,
\begin{eqnarray}\label{C4}
\frac{\partial P}{\partial \zeta}=2\zeta(\mu^{2}-m^{2})=0.
\end{eqnarray}
This implies that $\zeta$ is nonvanishing when $|\mu|=m$. Hence, using the above definitions/notations, the condensate $\zeta$ is given by
\begin{eqnarray}\label{C5}
\zeta^{2}=\frac{1}{2m}[n_{\text{tot}}-n_{\text{th}}(\mu=m)],
\end{eqnarray}
where $n_{\text{th}}\in\{n_{\text{nr}},n_{\text{ur}};n^{ (0)}_{\text{nr}},n^{ (0)}_{\text{ur}}\}$. In following sections, we compute $\zeta^{2}$ in each of the above cases.
%-----------------------------------------
\subsection{BEC in a nonrotating Bose gas in NR and UR limits}\label{sec4a}
%-----------------------------------------
Before presenting the results for a slowly rotating Bose gas in NR and UR limits, it is instructive to compute the condensate for a nonrotating Bose gas in these limits and determine the critical temperature $T_{c}$ of the BEC. The latter is characterized by the specific temperature at which $\zeta$ vanishes and $n_{\text{th}}=n_{\text{tot}}$.
\par
To this purpose, let us consider \eqref{C5} and insert $n_{\text{nr}}^{ (0)}(\mu=m)$ from \eqref{B17} into it. Having in mind that for $\mu=m$, the fugacity $z=1$ [see \eqref{B1}] and $g_{s}(z=1)=\zeta(s), \forall s$, we arrive at
\begin{eqnarray}\label{C6}
\zeta^{(0)2}_{\text{nr}}=\frac{1}{2m}\left(n_{\text{tot}}-\frac{1}{\lambda_{T}^{3}}\zeta\left(3/2\right)\right),
\end{eqnarray}
with the thermal length defined in \eqref{B9}. This leads immediately to the critical temperature in NR limit, arising from $n_{\text{tot}}\lambda_{T_c}^{3}=\zeta(3/2)$ at $T=T_{c}$,
\begin{eqnarray}\label{C7}
T_{c,\text{nr}}^{(0)}=\frac{2\pi}{m}\left(\frac{n_{\text{tot}}}{\zeta(3/2)}\right)^{2/3},
\end{eqnarray}
as expected from \cite{kapusta-book}. Using the fact that at $T=T_{c}$, $g_{3/2}(z=1)=\zeta(3/2)$, and at temperatures above the critical temperature $n_{\text{nr}}^{ (0)}=n_{\text{tot}}$, two important relations arise in NR limit
\begin{eqnarray}\label{C8}
n_{\text{tot}}\lambda_{T_c}^{3}&=&\zeta(3/2), \qquad \text{at $T=T_{c,\text{nr}}^{(0)}$},\nonumber\\
n_{\text{tot}}\lambda_{T}^{3}&=&g_{3/2}(z), \qquad \text{at $T>T_{c,\text{nr}}^{(0)}$},
\end{eqnarray}
where $\lambda_{T_c}\equiv \lambda_{T=T_{c,\text{nr}}^{(0)}}$. Combining these results, we thus obtain
\begin{eqnarray}\label{C9}
\frac{n_{0,\text{nr}}^{(0)}}{n_{\text{tot}}}&=&1-\left(\frac{T}{T_{c,\text{nr}}^{(0)}}\right)^{3/2},\nonumber\\
\frac{n_{\text{nr}}^{ (0)}}{n_{\text{tot}}}&=&\left(\frac{T}{T_{c,\text{nr}}^{(0)}}\right)^{3/2},\qquad \text{at $T\leq T_{c,\text{nr}}^{(0)}$}.
\end{eqnarray}
For a nonrotating Bose gas in UR limit, we remind that in this case the fugacity $\mathfrak{z}=e^{\beta\mu}$ [see \eqref{B19}]. For $\mu=m$ and in the limit $\beta m\to 0$ \cite{kapusta1979}, we have $\mathfrak{z}\simeq 1$. Using \eqref{C5} with $n_{\text{th}}=n_{\text{ur}}^{ (0)}$ from \eqref{B30}, the BE condensate in a nonrotating Bose gas in the UR limit is given by
\begin{eqnarray}\label{C10}
\zeta^{(0)2}_{\text{ur}}=\frac{1}{2m}\left(n_{\text{tot}}-\frac{1}{\Lambda_{T}^{3}}\zeta\left(3\right)\right),
\end{eqnarray}
with the thermal length $\Lambda_{T}$ from \eqref{B24}. Setting $\zeta^{(0)2}_{\text{ur}}=0$, we thus obtain $n_{\text{tot}}\lambda_{T_c}^{3}=\zeta(3)$. This determines the critical temperature of BEC in a nonrotating Bose gas in UR limit
\begin{eqnarray}\label{C11}
T_{c,\text{ur}}^{(0)}=\left(\frac{\pi^{2}n_{\text{tot}}}{\zeta(3)}\right)^{1/3}.
\end{eqnarray}
Moreover, we have
\begin{eqnarray}\label{C12}
n_{\text{tot}}\Lambda_{T_c}^{3}&=&\zeta(3), \qquad \text{at $T=T_{c,\text{ur}}^{(0)}$},\nonumber\\
n_{\text{tot}}\Lambda_{T}^{3}&=&g_{3}(\mathfrak{z}), \qquad \text{at $T>T_{c,\text{ur}}^{(0)}$},
\end{eqnarray}
and
\begin{eqnarray}\label{C13}
\frac{n_{0,\text{ur}}^{(0)}}{n_{\text{tot}}}&=&1-\left(\frac{T}{T_{c,\text{ur}}^{(0)}}\right)^{3},\nonumber\\
\frac{n_{\text{ur}}^{ (0)}}{n_{\text{tot}}}&=&\left(\frac{T}{T_{c,\text{ur}}^{(0)}}\right)^{3},\qquad \text{at $T\leq T_{c,\text{ur}}^{(0)}$}.
\end{eqnarray}
It is noteworthy that the expression \eqref{C11} for the critical temperature is different from that presented in \cite{kapusta-book}. To explain this difference, we remind that the above result arises only from the contribution of particles in the UR limit, while the result appearing in \cite{kapusta-book} arises by including the contribution of particles \textit{and} antiparticles to the number density $n$. In this case, the corresponding condensate $\zeta_{\text{ur}}^{(0)2}$ is first given by
\begin{eqnarray}\label{C14}
\zeta_{\text{ur}}^{(0)2}=\frac{1}{2m}\left\{n_{\text{tot}}-\frac{T^{3}}{\pi^{2}}\left[g_{3}(e^{\beta m})-g_{3}(e^{-\beta m})\right]\right\}. \nonumber\\
\end{eqnarray}
Taking the limit $\beta m\to 0$ and using
\begin{eqnarray}\label{C15}
g_{3}(e^{\pm \beta m})\approx \zeta(3)\pm \frac{\pi^{2}}{6}\beta m,
\end{eqnarray}
the condensate in UR limit reads
\begin{eqnarray}\label{C16}
\zeta_{\text{ur}}^{(0)2}=\frac{1}{2m}\left(n_{\text{tot}}-\frac{mT^{2}}{3}\right).
\end{eqnarray}
This implies immediately the critical temperature in UR limit
\begin{eqnarray}\label{C17}
T_{c}^{(0)}\big|_{\text{ur}}=\left(\frac{3 n_{\text{tot}}}{m}\right)^{1/2},
\end{eqnarray}
as presented in \cite{kapusta-book}.
\par
In what follows, we consider only the contribution of particles and determine the condensate and the BEC critical temperature in the NR and UR limits once $\Omega\neq 0$.
%-----------------------------------------
\subsection{BEC in nonrelativistic Bose gas with $\boldsymbol{\Omega\neq 0}$}\label{sec4b}
%-----------------------------------------
Let us consider $n_{\text{nr}}$ from \eqref{B14}. For $\mu=m$, the fugacity $z=1$, and $n_{\text{nr}}$ reduces to
\begin{eqnarray}\label{C18}
n_{\text{nr}}=\frac{1}{\lambda_{T,\Omega}^{3}}\zeta(5/2),
\end{eqnarray}
where $\lambda_{T,\Omega}=\lambda_{T}(\beta\Omega)^{1/3}$. Plugging this expression into \eqref{C5}, the BE condensate $\zeta^{2}_{\text{nr}}$ of a slowly rotating Bose gas in the NR limit is given by
\begin{eqnarray}\label{C19}
\zeta^{2}_{\text{nr}}=\frac{1}{2m}\left(n_{\text{tot}}-\frac{1}{\lambda_{T,\Omega}^{3}}\zeta(5/2)\right).
\end{eqnarray}
Using the fact that at BE critical temperature the condensate $\zeta^{2}_{\text{nr}}$ vanishes, we arrive at
\begin{eqnarray}\label{C20}
T_{c,\text{nr}}=\left(\frac{2\pi}{m}\right)^{3/5}\left(\frac{n_{\text{tot}}\Omega}{\zeta(5/2)}\right)^{2/5},
\end{eqnarray}
for fixed $n_{\text{tot}}$ and $\Omega$. This result is in contrast to \eqref{C7} for nonrotating case. According to this result, the critical temperature increases with increasing $\Omega$. Using the fact that at the critical temperature and above it, $n_{\text{nr}}=n_{\text{tot}}$, following relations arise in NR limit
\begin{eqnarray}\label{C21}
n_{\text{tot}}\lambda_{T_c,\Omega}^{3}&=&\zeta(5/2), \qquad \text{at $T=T_{c,\text{nr}}$}, \nonumber\\
n_{\text{tot}}\lambda_{T,\Omega}^{3}&=&g_{5/2}(z), \qquad \text{at $T>T_{c,\text{nr}}$}.
\end{eqnarray}
They replace \eqref{C8} for the nonrotating case. The condensate and thermal fractions are thus given by
\begin{eqnarray}\label{C22}
\frac{n_{0,\text{nr}}}{n_{\text{tot}}}&=&1-\left(\frac{T}{T_{c,\text{nr}}}\right)^{5/2},\nonumber\\
\frac{n_{\text{nr}}^{ }}{n_{\text{tot}}}&=&\left(\frac{T}{T_{c,\text{nr}}}\right)^{5/2},\qquad \text{at $T\leq T_{c,\text{nr}}$},
\end{eqnarray}
in contrast to the result for the nonrotating Bose gas \eqref{C9}. In addition, combining \eqref{C7} and \eqref{C20}, the critical temperature $T_{c,\text{nr}}$ and $T_{c\text{nr}}^{(0)}$ for a rotating and nonrotating Bose gas, and the fact that $n_{\text{tot}}=n_{\text{th}}(T=T_{c,\text{nr}}^{(0)})=n_{\text{th}}(T=T_{c,\text{nr}})$, we arrive first at
\begin{eqnarray}\label{C23}
T_{c,\text{nr}}=\left(\frac{\zeta(3/2)}{\zeta(5/2)}\right)^{2/5}\Omega^{2/5}\left(T_{c,\text{nr}}^{(0)}\right)^{3/5}.
\end{eqnarray}
Plugging this expression into \eqref{C22}, we then obtain
\begin{eqnarray}\label{C24}
\frac{n_{0,\text{nr}}}{n_{\text{tot}}}=1-\left(\frac{T}{T_{c,\text{nr}}^{(0)}}\right)^{3/2}\frac{\zeta(5/2)}{\zeta(3/2)}\frac{1}{\beta\Omega}.
\end{eqnarray}
This is the relation between $n_{0,\text{nr}}/n_{\text{tot}}$ to $T_{c,\text{nr}}^{(0)}$ and
$\Omega$.
\par
Apart from critical temperature, \eqref{C19} yields the critical angular velocity $\Omega_{c}$. Plugging the definition of $\lambda_{T,\Omega}$ into \eqref{C19}, $\zeta_{\text{nr}}=0$ leads to
\begin{eqnarray}\label{C25}
\Omega_{c,\text{nr}}=\left(\frac{m}{2\pi}\right)^{3/2}\frac{\zeta(5/2)}{n_{\text{tot}}}T^{5/2},
\end{eqnarray}
for fixed $n_{\text{tot}}$ and $T$. According to this result, $\Omega_{c,\text{nr}}$ increases with increasing temperature and is inversely proportional to $n_{\text{tot}}$. For $\Omega\leq \Omega_{c,\text{nr}}$, where $z=1$, it turns out that
\begin{eqnarray}\label{C26}
\hspace{-1cm}\frac{n_{0,\text{nr}}}{n_{\text{tot}}}=1-\left(\frac{\Omega}{\Omega_{c,\text{nr}}}\right)^{-1},\quad
\frac{n_{\text{nr}}}{n_{\text{tot}}}=\left(\frac{\Omega}{\Omega_{c,\text{nr}}}\right)^{-1}.
\end{eqnarray}
For $\Omega\geq\Omega_{c,\text{nr}}$, we particularly have $n_{\text{nr}}=n_{\text{tot}}$.
%-----------------------------------------
\subsection{BEC in ultrarelativistic Bose gas with $\boldsymbol{\Omega\neq 0}$}\label{sec4c}
%-----------------------------------------
\begin{figure*}[hbt]
\includegraphics[width=8cm, height=6cm]{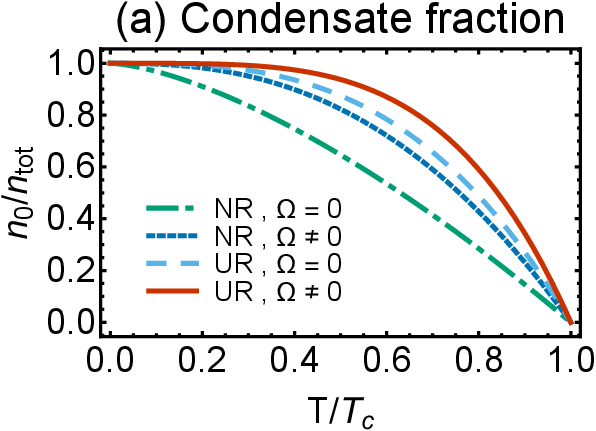}
\includegraphics[width=8cm, height=6cm]{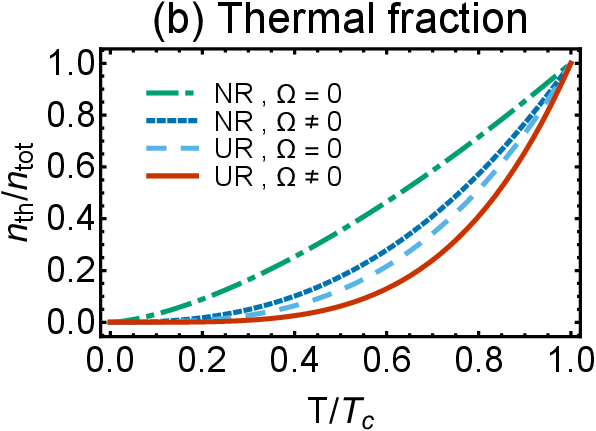}
\caption{color online. The $T/T_{c}$ dependence of the condensate fraction $n_{0}/n_{\text{tot}}$ (panel a) and thermal fraction $n_{\text{th}}/n_{\text{tot}}$ (panel b) is plotted for $n_{0}\in\{n_{0,\text{nr}}^{(0)}, n_{0,\text{ur}}^{(0)}\}$ (panel a) as well as $n_{\text{th}}\in\{n_{\text{nr}}^{(0)}, n_{\text{ur}}^{(0)}\}$ (panel b)
and $T_{c}\in\{T_{c,\text{nr}}^{(0)}, T_{c,\text{ur}}^{(0)}\}$ for a nonrotating Bose gas in NR and UR limits with $\Omega=0$ and $n_{0}\in\{n_{0,\text{nr}},n_{0,\text{ur}}\}$ (panel a) as well as $n_{\text{th}}\in\{n_{\text{nr}},n_{\text{ur}}\}$ (panel b)
 and $T_{c}\in\{T_{c,\text{nr}},T_{c,\text{ur}}\}$ for a rotating Bose gas in NR and UR limits. Nonrelativistic Bose gas under rigid rotation exhibits similar behavior as an ultrarelativistic Bose gas with $\Omega=0$.}\label{fig1}
\end{figure*}
\begin{figure*}[hbt]
\includegraphics[width=8cm, height=6cm]{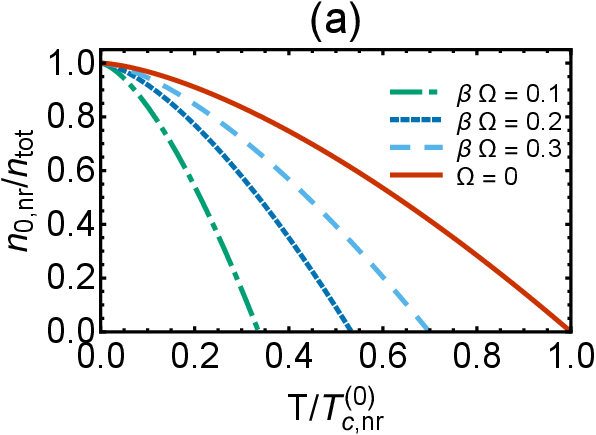}
\includegraphics[width=8cm, height=6cm]{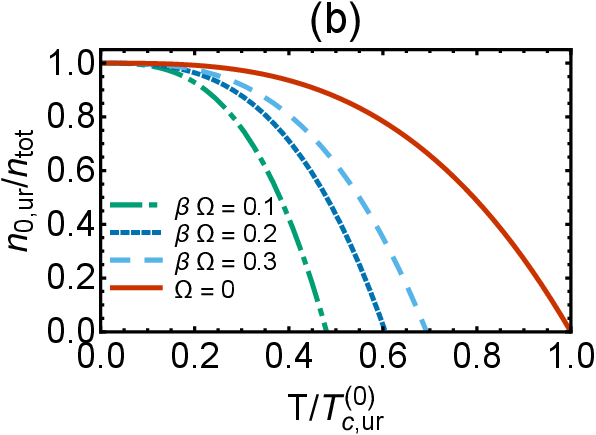}
\caption{color online. The $T/T_{c,\text{nr}}^{(0)}$ (panel a) and $T/T_{c,\text{ur}}^{(0)}$  (panel b) dependence of the condensate fraction $n_{0,\text{nr}}/n_{\text{tot}}$ (panel a) and $n_{0,\text{ur}}/n_{\text{tot}}$ (panel b) of rotating Bose gas in NR and UR limits are plotted for different $\beta\Omega=0.1,0.2,0.3$. Equations \eqref{C24} and \eqref{C33} are used. Red solid curves demonstrate the $T/T_{c}$ dependence of the condensate fraction $n_{0}^{(0)}/n_{\text{tot}}$ of nonrotating Bose gas in NR (panel a) and UR (panel b) limits. Here $n_{0}^{(0)}\in \{n_{0,\text{nr}}^{(0)},n_{0,\text{ur}}^{(0)}\}$ and $T_{c}^{(0)}\in \{T_{c,\text{nr}}^{(0)},T_{c,\text{ur}}^{(0)}\}$ for these two limits. Comparing with $T_{c,\text{nr}}^{(0)}$, the critical temperature of rotating Bose gas is lower, but increases with increasing $\beta\Omega$. }\label{fig2}
\end{figure*}
\begin{figure*}[hbt]
\includegraphics[width=8cm, height=6cm]{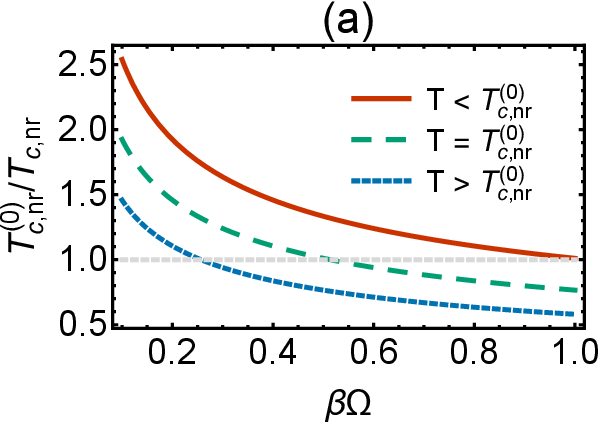}
\includegraphics[width=8cm, height=6cm]{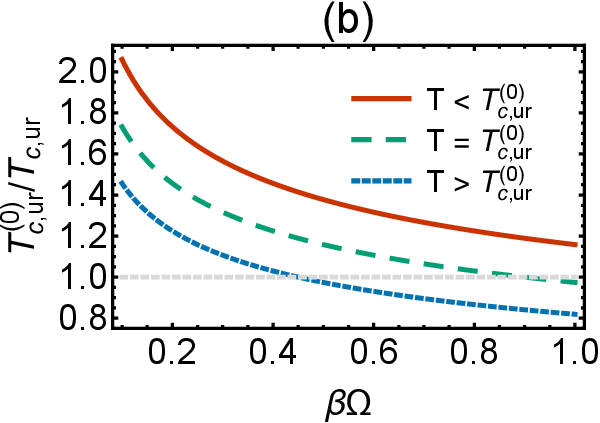}
\caption{color online. The $\beta\Omega$ dependence of $T_{c,\text{nr}}^{(0)}/T_{c,\text{nr}}$  (panel a) and $T_{c,\text{ur}}^{(0)}/T_{c,\text{ur}}$ (panel b) is plotted. Gray dashed lines indicate $T_{c,\text{nr}}^{(0)}=T_{c,\text{nr}}$ in panel (a) and $T_{c,\text{ur}}^{(0)}=T_{c,\text{ur}}$ in panel (b). According to these results, for $T>T_{c,\text{nr}}$ and $T>T_{c,\text{ur}}$, the critical temperature of nonrotating Bose gas in NR and UR limits is, independent of $\beta\Omega$, larger than the critical temperature of a rotating Bose gas in these two limits. For $T=T_{c}$ and $T<T_{c}$ with $T_{c}\in\{T_{c,\text{nr}},T_{c,\text{ur}}\}$, this statement is only valid for smaller values of $\beta\Omega$.}\label{fig3}
\end{figure*}
\begin{figure}[hbt]
\includegraphics[width=8cm, height=6cm]{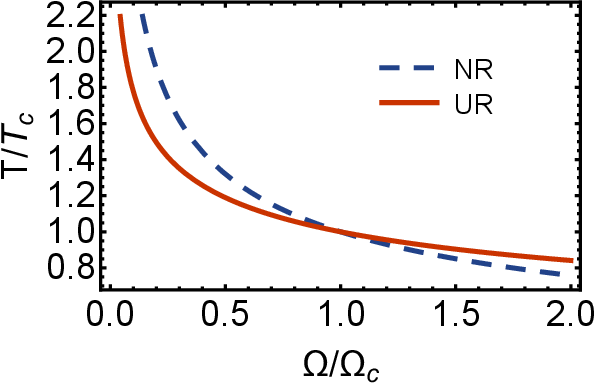}
\caption{color online. The $\Omega/\Omega_{c}$ dependence of the ratio $T/T_{c}$ is plotted for a nonrelativistic (NR) and an ultrarelativistic (UR) Bose gas under rigid rotation [see \eqref{C36} and \eqref{C37}]. Here, $T_{c}\in \{T_{c,\text{nr}}, T_{c,\text{ur}}\}$ and $\Omega_{c}\in \{\Omega_{c,\text{nr}}, \Omega_{c,\text{ur}}\}$.}\label{fig4}
\end{figure}
To determine the BE condensate fraction of a slowly rotating Bose gas in UR limit, we consider $n_{\text{ur}}$ from \eqref{B27}, and set $\mathfrak{z}=e^{\beta m}$ to satisfy the condition $\mu=m$ for the BE transition. Expanding $g_{4}(e^{\beta m})$ in the orders of $\beta m$, we arrive for $\beta m\ll 1$ at
\begin{eqnarray}\label{C27}
g_{4}(e^{\beta m})=\zeta(4)+\mathcal{O}(\beta m)\simeq \zeta(4),
\end{eqnarray}
where $\zeta(4)=\pi^{4}/90$. Plugging this expressions into \eqref{C6}, we obtain
\begin{eqnarray}\label{C28}
\zeta_{\text{ur}}^{2}=\frac{1}{2m}\left(n_{\text{tot}}-\frac{1}{\Lambda_{T,\Omega}^{3}}\zeta(4)\right),
\end{eqnarray}
where $\Lambda_{T,\Omega}=\Lambda_{T}(\beta \Omega)^{1/3}$. For $\zeta_{\text{ur}}^{2}=0$, the corresponding critical temperature of BE transition for fixed $n_{\text{tot}}$ and $\Omega$ reads
\begin{eqnarray}\label{C29}
T_{c,\text{ur}}=\left(\frac{\pi^{2}n_{\text{tot}}\Omega}{\zeta(4)}\right)^{1/4},
\end{eqnarray}
for fixed $\Omega$ and $n_{\text{tot}}$.
This result is different from \eqref{C11} for a nonrotating Bose gas in UR limit. Similar to previous cases, however, $n_{\text{ur}}=n_{\text{tot}}$ at the critical temperature. We thus have
\begin{eqnarray}\label{C30}
n_{\text{tot}}\Lambda_{T_c,\Omega}^{3}&=&\zeta(4),\qquad \text{at $T=T_{c,\text{ur}}$},\nonumber\\
n_{\text{tot}}\Lambda_{T,\Omega}^{3}&=&g_{4}\left(\mathfrak{z}\right),\quad~~ \text{at $T>T_{c,\text{ur}}$},
\end{eqnarray}
in contrast to \eqref{C12} in nonrotating Bose gas. Moreover, we obtain
\begin{eqnarray}\label{C31}
\frac{n_{0,\text{ur}}}{n_{\text{tot}}}&=&1-\left(\frac{T}{T_{c,\text{ur}}}\right)^{4},\nonumber\\
\frac{n_{\text{ur}}}{n_{\text{tot}}}&=&\left(\frac{T}{T_{c,\text{ur}}}\right)^{4},\qquad \text{at $T\leq T_{c,\text{ur}}$}.
\end{eqnarray}
Comparing \eqref{C11} and \eqref{C29}, it is possible to write $T_{c,\text{ur}}$ in terms of $T_{c,\text{ur}}^{(0)}$ as
\begin{eqnarray}\label{C32}
T_{c,\text{ur}}=\left(\frac{\zeta(3)}{\zeta(4)}\right)^{1/4}\Omega^{1/4}\left(T_{c,\text{ur}}^{(0)}\right)^{3/4}.
\end{eqnarray}
Plugging \eqref{C32} into \eqref{C31},
 $n_{0,\text{ur}}/n_{\text{tot}}$ can be expressed in terms of  $T_{c,\text{ur}}^{(0)}$,
\begin{eqnarray}\label{C33}
\frac{n_{0,\text{ur}}}{n_{\text{tot}}}&=&1-\left(\frac{T}{T_{c,\text{ur}}^{(0)}}\right)^{3}\frac{\zeta(4)}{\zeta(3)}\frac{1}{\beta\Omega}.
\end{eqnarray}
Similar to the previous case, $\zeta_{\text{ur}}^{2}=0$ leads also to a critical angular velocity $\Omega_{c,\text{ur}}$ in UR limit for fixed $n_{\text{tot}}$ and $T$,
\begin{eqnarray}\label{C34}
\Omega_{c,\text{ur}}=\frac{T^{4}\zeta(4)}{\pi^{2}n_{\text{tot}}}.
\end{eqnarray}
Combining this result with \eqref{C31} yields for $\Omega\leq \Omega_{c,\text{ur}}$,
\begin{eqnarray}\label{C35}
\hspace{-1cm}\frac{n_{0,\text{ur}}}{n_{\text{tot}}}=1-\left(\frac{\Omega}{\Omega_{c,\text{ur}}}\right)^{-1},\quad
\frac{n_{\text{ur}}}{n_{\text{tot}}}=\left(\frac{\Omega}{\Omega_{c,\text{ur}}}\right)^{-1}.
\end{eqnarray}
In Fig. \ref{fig1}, the $T/T_{c}$ dependence of the  condensate and thermal fractions $n_{0}/n_{\text{tot}}$ and $n_{\text{th}}/n_{\text{tot}}$ [see Fig. \ref{fig1}(a) and \ref{fig1}(b)] are plotted for nonrotating and rotating Bose gas in NR and UR limits. For nonrotating gas in NR and UR  limits, $n_{0,\text{nr}}^{(0)}/n_{\text{tot}}$ and $n_{\text{nr}}^{(0)}/n_{\text{tot}}$ as well as $n_{0,\text{ur}}^{(0)}/n_{\text{tot}}$ and $n_{\text{ur}}^{(0)}/n_{\text{tot}}$ from \eqref{C9} and \eqref{C13} are used. For rotating Bose gas in NR and UR limits,  $n_{0,\text{nr}}/n_{\text{tot}}$ and $n_{\text{nr}}/n_{\text{tot}}$ as well as $n_{0,\text{ur}}/n_{\text{tot}}$ and $n_{\text{ur}}/n_{\text{tot}}$ from \eqref{C22} and \eqref{C31} are used. The difference between the slope of the curves arises from the exponents of $T/T_{c}$ appearing in the above expressions. According to these results, a nonrelativistic Bose gas under rigid rotation exhibits similar behavior as a nonrotating gas in UR limit.
\par
In order to explore the effect of $\Omega$ on the production rates of the BE condensate, the $T/T_{c}^{(0)}$ dependence of $n_{0}/n_{\text{tot}}$ is plotted in Fig. \ref{fig2}. In Fig. \ref{fig2}(a) $n_{0}\in\{n_{0,\text{nr}}^{(0)},n_{0,\text{nr}}\}$ corresponds to the number density of condensates in nonrotating and rotating gases in NR limit and in Fig. \ref{fig2}(b) $n_{0}\in\{n_{0,\text{ur}}^{(0)},n_{0,\text{ur}}\}$ corresponding to the same quantity in nonrotating and rotating gases in UR limits. To plot the curves in Fig. \ref{fig2}(a) and \ref{fig2}(b), we used \eqref{C24} and \eqref{C33}, respectively. According to these results, a comparison between $T_{c,\text{nr}}^{(0)}$ as well as $T_{c,\text{ur}}^{(0)}$ with the corresponding critical temperatures of rotating gases in NR and UR limits shows that the critical temperatures of BE transition decreases once the Bose gas is subjected to a rigid rotation. This would mean that a bosonic gas becomes warmer as it starts to rotate. However, once the critical temperatures of \textit{rotating} gases are compared with each other, they increase with increasing $\beta \Omega$. In other words, increasing the angular velocity lowers the temperature of a rotating Bose gas. Moreover, for a constant $T/T_{c,\text{nr}}$ or $T/T_{c,\text{ur}}$, the condensate fraction increases with increasing $\beta\Omega$, as expected from \eqref{C24} and \eqref{C33}.
\par
Using \eqref{C23} and \eqref{C32}, it is possible to determine the ratio $T_{c,\text{nr}}^{(0)}/T_{c,\text{nr}}$ and $T_{c,\text{ur}}^{(0)}/T_{c,\text{ur}}$ as a function of $\beta\Omega$ and $T$. In Fig. \ref{fig3}, the $\beta\Omega$ dependence of this ratio is plotted [see Figs. \ref{fig3}(a) and \ref{fig3}(b), for nonrelativistic and ultrarelativistic gases]. According to these results, for $T>T_{c,\text{nr}}^{(0)}$ as well as $T>T_{c,\text{ur}}^{(0)}$, the critical temperature of a nonrotating gas is, independent of $\beta\Omega$, always larger than the critical temperature of a rotating gas. At $T\geq T_{c}^{(0)}$, however, the decrease of the critical temperature caused by rotation depends on $\beta\Omega$. At $T=T_{c,\text{nr}}^{(0)}$, e.g., for $\beta\Omega\lesssim 0.5$, $T_{c,\text{nr}}^{(0)}>T_{c,\text{nr}}$, while for $\beta\Omega>0.5$ we have $T_{c,\text{nr}}^{(0)}>T_{c,\text{nr}}$ [$\beta\Omega\simeq 0.5$ arises from the intersection of green dashed curve with the horizontal gray dashed line in Fig. \ref{fig3}(a)]. For $T>T_{c,\text{nr}}^{(0)}$, the $\beta\Omega$ arising from the intersection of blue dashed curve with the horizontal gray dashed curve is even smaller. The same is also true for ultrarelativistic Bose gas. Let us remind that analytical results presented in this paper, are principally valid only for $\beta\Omega\ll 1$. In this case, the critical temperature of rotating Bose gas is always smaller than the critical temperature of a nonrotating one, no matter whether the gas is non- or ultrarelativistic.
\par
In Fig. \ref{fig4}, the $\Omega/\Omega_{c}$ dependence of $T/T_{c}$ is plotted for rotating Bose gas in NR (blue dashed curve) and UR (red solid curve) limits. This relation is determined by combining \eqref{C22} and \eqref{C26} for rotating Bose gas in NR limit as well as \eqref{C31} and \eqref{C35} for rotating gas in UR limit. The former one leads to
\begin{eqnarray}\label{C36}
\frac{T}{T_{c,\text{nr}}}=\left(\frac{\Omega}{\Omega_{c,\text{nr}}}\right)^{-2/5},
\end{eqnarray}
 for a rotating gas in NR limit and the latter one to
\begin{eqnarray}\label{C37}
\frac{T}{T_{c,\text{ur}}}=\left(\frac{\Omega}{\Omega_{c,\text{ur}}}\right)^{-1/4},
\end{eqnarray}
for a rotating gas in UR limit. Clearly, at $T=T_{c}$, the system is automatically at $\Omega=\Omega_{c}$, independent of chosen limit.
\par
In the next section, we use the above analytical results to determine the values of various thermodynamical quantities below and above the BE transition temperatures for nonrotating and rotating gases in NR and UR limits.
%-----------------------------------------
\section{Thermodynamic properties of Bose gas below and above the BEC critical temperature}\label{sec5}
\setcounter{equation}{0}
%-----------------------------------------
In this section, we study the behavior of various thermodynamic quantities below and above the BEC critical temperature. To this purpose, we first determine the $T$ dependence of fugacity above $T_{c}$. We then reconsider the pressure $P$ and show that its behavior below and above $T_{c}$ is different. The same is also true for the heat capacity $C_{V}$. We also determine the entropy density at $T_{c}$. Combining the entropy and number density, we compute the corresponding latent heat to nonrotating and rotating gases in NR and UR limits.
%-----------------------------------------
\subsection{The fugacity at $\boldsymbol{T\leq T_{c}}$ and $\boldsymbol{T>T_{c}}$}\label{sec5a}
%-----------------------------------------
\subsubsection{Nonrotating Bose gas in NR and UR limits}\label{sec5a1}
%-----------------------------------------
As we have described in previous sections, the definition of fugacity in NR and UR limits are different. In NR limit, it is defined by $z=e^{\beta(\mu-m)}$ [see \eqref{B1}] and in UR limit by $\mathfrak{z}=e^{\beta\mu}$ [see \eqref{B19}]. As aforementioned, at temperatures below the critical temperature, the condition for building the condensate is $\mu=m$. In NR limit, this leads to $z=1$ and in UR limit $\mathfrak{z}=e^{\beta m}\simeq 1$, upon taking the limit $\beta m\to 0$. We thus have
\begin{eqnarray}\label{D1}
z=1&\qquad&\text{at $T\leq T_{c,\text{nr}}^{(0)}$},\nonumber\\
\mathfrak{z}\simeq 1&\qquad&\text{at $T\leq T_{c,\text{ur}}^{(0)}$}.
\end{eqnarray}
At temperatures higher than the critical temperature, the situation is different. In a nonrotating Bose gas the fugacity depends only on $T$, while in a rotating gas it depends on $T$ and $\Omega$. To determine the $T$ dependence of $z$ of a nonrotating gas in NR limit, let us consider \eqref{C8}. Having in mind that $z=e^{-\alpha}$ with $\alpha\equiv -\beta(\mu-m)$, we evaluate $g_{3/2}(e^{-\alpha})$ for $\alpha\to 0$ by making use of
\begin{eqnarray}\label{D2}
g_{\nu}(e^{-\alpha})=\frac{\Gamma(1-\nu)}{\alpha^{1-\nu}}+\sum_{j=0}^{\infty}\frac{(-1)^{j}}{j!}\zeta(\nu-j)\alpha^{j}.
\end{eqnarray}
We arrive first at
\begin{eqnarray}\label{D3}
g_{3/2}(z)\simeq -(4\pi\alpha)^{1/2}+\zeta(3/2),
\end{eqnarray}
leading to
\begin{eqnarray}\label{D4}
\alpha\simeq \frac{1}{4\pi}(\zeta(3/2))^{2}\left(1-\frac{g_{3/2}(z)}{\zeta(3/2)}\right)^{2}.
\end{eqnarray}
Plugging at this stage
\begin{eqnarray}\label{D5}
\frac{g_{3/2}(z)}{\zeta(3/2)}=\left(\frac{T_{c,\text{nr}}^{(0)}}{T}\right)^{3/2},
\end{eqnarray}
arising from \eqref{C8}, into \eqref{D4} and using
\begin{eqnarray}\label{D6}
(1-x^{-3/2})\xrightarrow{x\to 1}\frac{3}{2}(x-1),
\end{eqnarray}
with $x\equiv T/T_{c,\text{nr}}^{(0)}$, we obtain
\begin{eqnarray}\label{D7}
\alpha\simeq \frac{9}{16\pi}(\zeta(3/2))^{2}\left(1- \frac{T}{T_{c,\text{nr}}^{(0)}}\right)^{2}.
\end{eqnarray}
This leads to the fugacity of a nonrotating Bose gas in NR limit
\begin{eqnarray}\label{D8}
z\simeq \exp\left(-\frac{9}{16\pi}(\zeta(3/2))^{2}\left(\frac{T}{T_{c,\text{nr}}^{(0)}}-1\right)^{2}\right),\qquad \text{at $T> T_{c,\text{nr}}^{(0)}$}. \nonumber\\
\end{eqnarray}
Let us notice that \eqref{C8} yields another useful relation. Differentiating both sides of \eqref{C8} with respect to $T$ and using
\begin{eqnarray}\label{D9}
\frac{\partial g_{\nu}(z)}{\partial z}=\frac{1}{z}g_{\nu-1}(z),
\end{eqnarray}
we obtain
\begin{eqnarray}\label{D10}
\left(\frac{\partial z}{\partial T}\right)_{n}=-\frac{3z}{2T}\frac{g_{3/2}(z)}{g_{1/2}(z)},\qquad \text{at $T> T_{c,\text{nr}}^{(0)}$}.
\end{eqnarray}
Following the same procedure, it is possible to derive similar expression for the fugacity $\mathfrak{z}$ for a nonrotating gas in UR limit. In this case, we use \eqref{C12} and
\begin{eqnarray}\label{D11}
g_{3}(\mathfrak{z})\simeq \zeta(3)+\mathfrak{a}\zeta(2),
\end{eqnarray}
with $\mathfrak{a}\equiv \ln \mathfrak{z}=\beta \mu$, to arrive first at
\begin{eqnarray}\label{D12}
\mathfrak{a}\simeq\frac{\zeta(3)}{\zeta(2)}\left(\frac{g_{3}(\mathfrak{z})}{\zeta(3)}-1\right).
\end{eqnarray}
Then, combining the relations from \eqref{C12}, we obtain
\begin{eqnarray}\label{D13}
\mathfrak{a}\simeq\frac{\zeta(3)}{\zeta(2)}\left(\left(\frac{T_{c,\text{ur}}^{(0)}}{T}\right)^{3}-1\right).
\end{eqnarray}
Using
\begin{eqnarray}\label{D14}
(x^{-3}-1)\xrightarrow{x\to 1}-3\left(x-1\right),
\end{eqnarray}
with $x\equiv  T/T_{c,\text{ur}}^{(0)}$, we arrive at the fugacity of a nonrotating Bose gas in UR limit
\begin{eqnarray}\label{D15}
\mathfrak{z}\simeq \exp\left(-3\frac{\zeta(3)}{\zeta(2)}\left( \frac{T}{T_{c,\text{ur}}^{(0)}}-1\right)\right), \qquad \text{at $T> T_{c,\text{ur}}^{(0)}$}.\nonumber\\
\end{eqnarray}
Differentiating \eqref{C12} with respect to $T$ and using \eqref{D9} results in
\begin{eqnarray}\label{D16}
\left(\frac{\partial \mathfrak{z}}{\partial T}\right)_{n}=-\frac{3\mathfrak{z}}{T}\frac{g_{3}(\mathfrak{z})}{g_{2}(\mathfrak{z})},\qquad \text{at $T> T_{c,\text{ur}}^{(0)}$}.
\end{eqnarray}
We use \eqref{D10} and \eqref{D16} in Sec. \ref{sec5c} to determine the heat capacity of nonrotating Bose gas at temperatures above the critical temperature.
%-----------------------------------------
\subsubsection{Rotating Bose gas in NR and UR limits}\label{sec5a2}
%-----------------------------------------
\begin{figure*}[hbt]
\includegraphics[width=8cm, height=6cm]{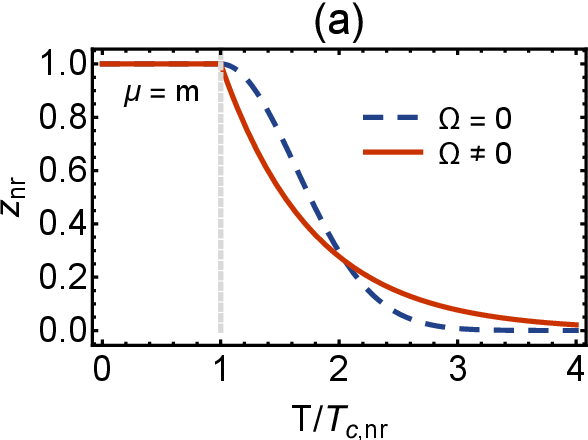}
\includegraphics[width=8cm, height=6cm]{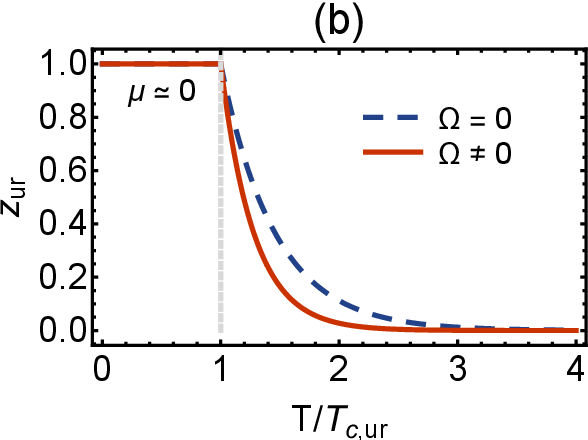}
\caption{color online. The $T/T_{c}$ dependence of the fugacity is plotted for nonrelativistic (panel a) and ultrarelativistic (panel b) Bose gases in the absence (blue dashed curves) and presence (red solid curves) of rigid rotation. Here, $T_{c}\in \{T_{c,\text{nr}}^{(0)}, T_{c,\text{ur}}^{(0)}\}$ for $\Omega=0$ and $T_{c}\in \{T_{c,\text{nr}}, T_{c,\text{ur}}\}$ for $\Omega\neq 0$. In contrast to the main text, the notion $z_{\text{nr}}=z$ and $z_{\text{ur}}=\mathfrak{z}$ is used.}\label{fig5}
\end{figure*}
Following the above procedure, we derive in this section the fugacity of a rotating Bose gas in NR and UR limits. Similar to a nonrotating Bose gas, the fugacity is equal to $1$ below the critical temperature,
\begin{eqnarray}\label{D17}
z=1&\qquad&\text{at $T\leq T_{c,\text{nr}}$},\nonumber\\
\mathfrak{z}\simeq 1&\qquad&\text{at $T\leq T_{c,\text{ur}}$}.
\end{eqnarray}
At temperatures higher than $T_{c}$, however, the fugacity depends on $T$ and $\Omega$, in contrast to the case of nonrotating Bose gas. To determine its dependence at $T> T_{c,\text{nr}}$ for a rotating Bose gas in NR limit, let us consider \eqref{C21} and use
\begin{eqnarray}\label{D18}
g_{5/2}(e^{-\alpha})\simeq \zeta(5/2)-\alpha\zeta(3/2),
\end{eqnarray}
and
\begin{eqnarray}\label{D19}
\frac{g_{5/2}(z)}{\zeta(5/2)}=\left(1-\left(\frac{T_{c,\text{nr}}}{T}\right)^{5/2}\right).
\end{eqnarray}
We thus arrive at
\begin{eqnarray}\label{D20}
z\simeq\exp\left(-\frac{5}{2}\frac{\zeta(5/2)}{\zeta(3/2)}\left(\frac{T}{T_{c,\text{nr}}}-1\right)\right),\qquad \text{at $T> T_{c,\text{nr}}$}, \nonumber\\
\end{eqnarray}
where
\begin{eqnarray}\label{D21}
(1-x^{-1})^{5/2}\xrightarrow{x\to 1}\frac{5}{2}\left(1-x\right),
\end{eqnarray}
is also used. Moreover, differentiating \eqref{C21} with respect to $T$ and $\Omega$ yields
\begin{eqnarray}\label{D22}
\left(\frac{\partial z}{\partial T}\right)_{n,\Omega}&=&-\frac{5z}{2T}\frac{g_{5/2}(z)}{g_{3/2}(z)},\nonumber\\
\left(\frac{\partial z}{\partial\Omega}\right)_{n,T}&=&\frac{z}{\Omega}\frac{g_{5/2}(z)}{g_{3/2}(z)},\qquad \text{at $T> T_{c,\text{nr}}$}.
\end{eqnarray}
For a rotating Bose gas in UR limit, we use \eqref{C30} and
\begin{eqnarray}\label{D23}
g_{4}(\mathfrak{z})\simeq \zeta(4)+\mathfrak{a}\zeta(3),
\end{eqnarray}
with $\mathfrak{a}=\ln\mathfrak{z}=\beta\mu$, to arrive first at
\begin{eqnarray}\label{D24}
\mathfrak{a}\simeq\frac{\zeta(4)}{\zeta(3)}\left(\frac{g_{4}(\mathfrak{z})}{\zeta(4)}-1\right).
\end{eqnarray}
The combination of relations from \eqref{C30} leads to
\begin{eqnarray}\label{D25}
\mathfrak{a}\simeq\frac{\zeta(4)}{\zeta(3)}\left(\left(\frac{T_{c,\text{ur}}^{(0)}}{T}\right)^{4}-1\right).
\end{eqnarray}
Utilizing
\begin{eqnarray}\label{D26}
(x^{-4}-1)\xrightarrow{x\to 1}-4\left(x-1\right),
\end{eqnarray}
the fugacity of a rotating Bose gas in UR limit reads
\begin{eqnarray}\label{D27}
\mathfrak{z}\simeq \exp\left(-4\frac{\zeta(4)}{\zeta(3)}\left( \frac{T}{T_{c,\text{ur}}}-1\right)\right), \qquad \text{at $T> T_{c,\text{ur}}$}.\nonumber\\
\end{eqnarray}
Differentiating \eqref{C30} with respect to $T$ and $\Omega$ leads to
\begin{eqnarray}\label{D28}
\left(\frac{\partial \mathfrak{z}}{\partial T}\right)_{n,\Omega}&=&-\frac{4\mathfrak{z}}{T}\frac{g_{4}(\mathfrak{z})}{g_{3}(\mathfrak{z})},\nonumber\\
\left(\frac{\partial \mathfrak{z}}{\partial\Omega}\right)_{n,T}&=&\frac{\mathfrak{z}}{\Omega}\frac{g_{4}(\mathfrak{z})}{g_{3}(\mathfrak{z})},\qquad \text{at $T> T_{c,\text{ur}}$}.
\end{eqnarray}
We use \eqref{D22} and \eqref{D28} in Sec. \ref{sec5c} to determine $C_{V}$ at $T>T_{c}$.
\par
In Fig. \ref{fig5}, the $T/T_{c}$ dependence of the fugacity $z_{\text{nr}}\equiv z$ and $z_{\text{ur}}\equiv\mathfrak{z}$ for nonrelativistic and ultrarelativistic Bose gas is plotted. Dashed blue (Solid red) curves correspond to nonrotating (rotating) Bose gases. We used \eqref{D1}, \eqref{D8}, and \eqref{D15} for a nonrotating as well as \eqref{D17}, \eqref{D20}, and \eqref{D27} for a rotating gas. According to these results, fugacities for temperatures below the critical temperatures are constant. They decrease with increasing temperature. The slope of these curves depends on whether the Bose gas is in NR or UR limit, and whether $\Omega =0$ or $\Omega\neq 0$.
%-----------------------------------------
\subsection{The pressure at $\boldsymbol{T\leq T_{c}}$ and $\boldsymbol{T>T_{c}}$}\label{sec5b}
%-----------------------------------------
\subsubsection{Nonrotating Bose gas in NR and UR limits}\label{sec5b1}
%-----------------------------------------
\begin{figure*}[hbt]
\includegraphics[width=8cm, height=6cm]{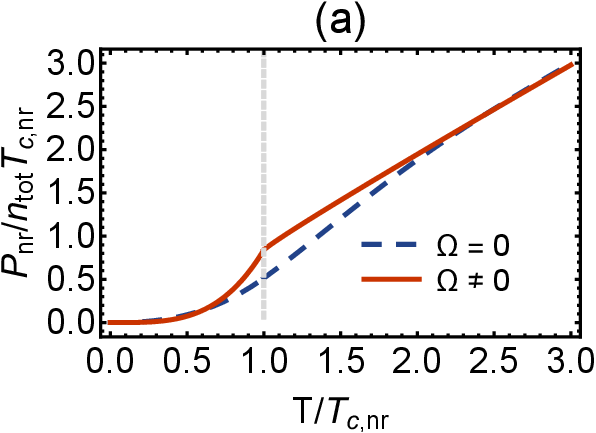}
\includegraphics[width=8cm, height=6cm]{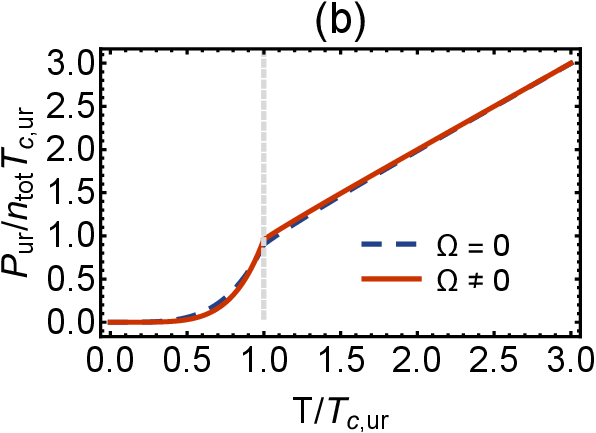}
\caption{color online. The $T/T_{c}$ dependence of dimensionless pressure $P/n_{\text{tot}}T_{c}$ is plotted for nonrelativistic (panel a) and ultrarelativistic (panel b) Bose gas in the absence (blue dashed curves) and presence (red solid curves) of rigid rotation.  Vertical gray dashed line indicates the position of $T=T_{c}$ with $T_{c}\in\{T_{c,\text{nr}}^{(0)},T_{c,\text{ur}}^{(0)}\}$ for nonrotating gas in NR and UR limits and $T_{c}\in\{T_{c,\text{nr}},T_{c,\text{ur}}\}$ for rotating Bose gas in these two limits. In all the above cases, the curves are continuous at $T=T_{c}$. }\label{fig6}
\end{figure*}
In Sec. \ref{sec3a}, we determined the thermal parts of the pressure $P_{\text{nr}}^{ (0)}$ and $P_{\text{ur}}^{ (0)}$ for a nonrotating Bose gas in NR and UR limits [see \eqref{B16} and \eqref{B29}]. In this section, we reconsider these expressions and using the results presented in Sec. \ref{sec4}, we compute these pressures at temperatures below and above the corresponding critical temperatures.
\par
To begin, let us consider the pressure of a nonrotating Bose gas in NR limit $P_{\text{nr}}^{ (0)}$ from \eqref{B16}. Using \eqref{C8}, we arrive at
\begin{eqnarray}\label{D29}
P_{\text{nr}}^{ (0)}&=&n_{\text{tot}}T_{c,\text{nr}}^{(0)}~\frac{\zeta(5/2)}{\zeta(3/2)}\left(\frac{T}{T_{c,\text{nr}}^{(0)}}\right)^{5/2},\qquad \text{at $T<T_{c,\text{nr}}^{(0)}$},\nonumber\\
P_{\text{nr}}^{ (0)}&=&n_{\text{tot}}T_{c,\text{nr}}^{(0)}~\frac{\zeta(5/2)}{\zeta(3/2)},\hspace{2.45cm}\text{at $T=T_{c,\text{nr}}^{(0)}$},\nonumber\\
P_{\text{nr}}^{ (0)}&=&n_{\text{tot}}T_{c,\text{nr}}^{(0)}~\frac{g_{5/2}(z)}{g_{3/2}(z)}\frac{T}{T_{c,\text{nr}}^{(0)}},\hspace{1.65cm} \text{at $T>T_{c,\text{nr}}^{(0)}$}.\nonumber\\
\end{eqnarray}
Here, $z$ is given in \eqref{D8}. Similarly, combining \eqref{B29} and \eqref{C12} leads to
\begin{eqnarray}\label{D30}
P_{\text{ur}}^{ (0)}&=&n_{\text{tot}}T_{c,\text{ur}}^{(0)}~\frac{\zeta(4)}{\zeta(3)}\left(\frac{T}{T_{c,\text{ur}}^{(0)}}\right)^{4},\qquad \text{at $T<T_{c,\text{ur}}^{(0)}$},\nonumber\\
P_{\text{ur}}^{ (0)}&=&n_{\text{tot}}T_{c,\text{ur}}^{(0)}~\frac{\zeta(4)}{\zeta(3)},\hspace{2.22cm}\text{at $T=T_{c,\text{ur}}^{(0)}$},\nonumber\\
P_{\text{ur}}^{ (0)}&=&n_{\text{tot}}T_{c,\text{ur}}^{(0)}~\frac{g_{4}(\mathfrak{z})}{g_{3}(\mathfrak{z})}\frac{T}{T_{c,\text{nr}}^{(0)}},\hspace{1.38cm} \text{at $T>T_{c,\text{ur}}^{(0)}$},\nonumber\\
\end{eqnarray}
where the $T$ dependence of $\mathfrak{z}$ at $T>T_{c,\text{ur}}^{(0)}$ is given in \eqref{D15}.
%-----------------------------------------
\subsubsection{Rotating Bose gas in NR and UR limits}\label{sec5b2}
%-----------------------------------------
In Sec. \ref{sec3a}, we compute the pressure $P$ for a rotating Bose gas in NR and UR limits in terms of the corresponding fugacities [see \eqref{B12} and \eqref{B26}]. Similar to the case of a nonrotating Bose gas in NR limit, we use \eqref{B12} and \eqref{C21} to arrive at the thermal pressure of a rotating gas in this limit
\begin{eqnarray}\label{D31}
P_{\text{nr}}^{ }&=&n_{\text{tot}}T_{c,\text{nr}}~\frac{\zeta(7/2)}{\zeta(5/2)}\left(\frac{T}{T_{c,\text{nr}}}\right)^{7/2},\qquad \text{at $T<T_{c,\text{nr}}$},\nonumber\\
P_{\text{nr}}^{ }&=&n_{\text{tot}}T_{c,\text{nr}}~\frac{\zeta(7/2)}{\zeta(5/2)},\hspace{2.45cm}\text{at $T=T_{c,\text{nr}}$},\nonumber\\
P_{\text{nr}}^{ }&=&n_{\text{tot}}T_{c,\text{nr}}~\frac{g_{7/2}(z)}{g_{5/2}(z)}\frac{T}{T_{c,\text{nr}}},\hspace{1.65cm} \text{at $T>T_{c,\text{nr}}$}.\nonumber\\
\end{eqnarray}
This is in contrast to the results \eqref{D29}. In this case, $z$ is $T$ and $\Omega$ dependent [see \eqref{D20}].
\begin{figure}[hbt]
\includegraphics[width=8cm, height=6cm]{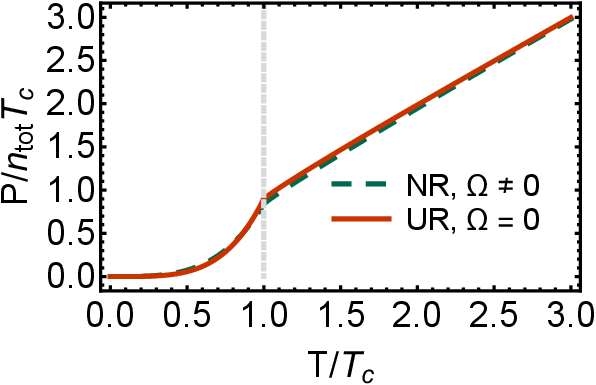}
\caption{color online. A comparison between the $T/T_{c,\text{nr}}$ dependence of dimensionless pressure $P_{\text{nr}}/n_{\text{tot}}T_{c,\text{nr}}$ corresponding to a rotating Bose gas in NR limit (dashed curve) with the $T/T_{c,\text{ur}}^{(0)}$ dependence of dimensionless pressure $P_{\text{ur}}^{(0)}/n_{\text{tot}}T_{c,\text{ur}}^{(0)}$ of a nonrotating gas in the UR limit (red solid curve). The fact that these curves almost overlap indicates that a nonrelativistic Bose gas under rotation behaves as a nonrotating gas in UR limit. }\label{fig7}
\end{figure}
\begin{figure*}[hbt]
\includegraphics[width=8cm, height=6cm]{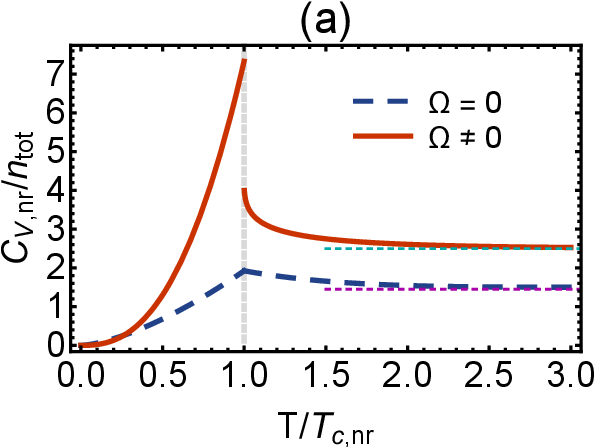}
\includegraphics[width=8cm, height=6cm]{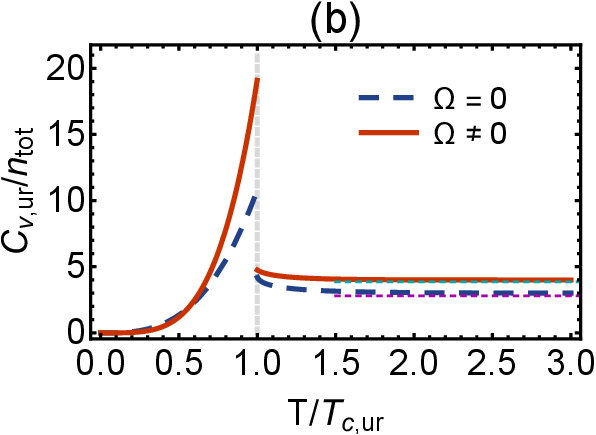}
\caption{color online. The $T/T_{c}$ dependence of dimensionless heat capacity $C_{V}/n_{\text{tot}}$ is plotted for a nonrelativistic (panel a) and an ultrarelativistic (panel b) Bose gas in the absence (blue dashed curves) and presence (red solid curves) of rigid rotation. The vertical gray dashed line indicates the position of $T=T_{c}$ with $T_{c}\in\{T_{c,\text{nr}}^{(0)},T_{c,\text{ur}}^{(0)}\}$ for nonrotating gas in NR and UR limits and $T_{c}\in\{T_{c,\text{nr}},T_{c,\text{ur}}\}$ for rotating Bose gas in these two limits. Apart from the case of nonrelativistic Bose gas in the absence of rotation, $C_{V}/n_{\text{tot}}$ are discontinuous at temperatures below and above the critical temperatures. In panel (a), curves asymptotically approach $C_{V,\text{nr}}^{(0)}/n_{\text{tot}}=3/2$ (magenta horizontal line) and $C_{V,\text{nr}}/n_{\text{tot}}=5/2$ (cyan horizontal line) at high temperatures $z\ll 1$.
In panel (b), curves asymptotically approach $C_{V,\text{ur}}^{(0)}/n_{\text{tot}}=3$ (magenta horizontal line) and $C_{V,\text{ur}}/n_{\text{tot}}=4$ (cyan horizontal line) at high temperatures $z\ll 1$.}\label{fig8}
\end{figure*}
Similarly, \eqref{B26} can be combined with \eqref{C30} to yield
\begin{eqnarray}\label{D32}
P_{\text{ur}}&=&n_{\text{tot}}T_{c,\text{ur}}~\frac{\zeta(5)}{\zeta(4)}\left(\frac{T}{T_{c,\text{ur}}}\right)^{5},\qquad \text{at $T<T_{c,\text{ur}}$},\nonumber\\
P_{\text{ur}}&=&n_{\text{tot}}T_{c,\text{ur}}~\frac{\zeta(5)}{\zeta(4)},\hspace{2.22cm}\text{at $T=T_{c,\text{ur}}$},\nonumber\\
P_{\text{ur}}&=&n_{\text{tot}}T_{c,\text{ur}}~\frac{g_{5}(\mathfrak{z})}{g_{4}(\mathfrak{z})}\frac{T}{T_{c,\text{ur}}},\hspace{1.38cm} \text{at $T>T_{c,\text{ur}}$},\nonumber\\
\end{eqnarray}
where in this case, the $T$ and $\Omega$ dependence of $\mathfrak{z}$ is given in \eqref{D27}. In Fig. \ref{fig6},  the $T/T_{c}$ dependence of dimensionless pressure $P/n_{\text{tot}}T_{c}$ is plotted for nonrelativistic [Fig. \ref{fig6}(a)] and ultrarelativistic [Fig. \ref{fig6}(b)] Bose gas for $\Omega =0$ (blue dashed curves) and $\Omega\neq 0 $ (red solid curves). Vertical gray dashed lines indicate the position of critical temperatures. According to the results \eqref{D29} and \eqref{D30}, $P_{\text{nr}}^{(0)}$ and $P_{\text{ur}}^{(0)}$ are continuous at $T=T_{c,\text{nr}}^{(0)}$ and $T=T_{c,\text{ur}}^{(0)}$. The same is also true for $P_{\text{nr}}$ and $P_{\text{ur}}$ from \eqref{D31} and \eqref{D32}. A comparison between the curves demonstrated in Fig. \ref{fig6}(a) show however that nonrelativistic gases in the absence and presence of rotation behave differently around the critical point. The specific form of $P_{\text{nr}}$ and
the nonanalyticity at $T=T_{c,\text{nr}}$ indicates that the heat capacity $C_{V,\text{nr}}$ is discontinuous at this temperature (see below). The same is also true for $P_{\text{ur}}^{(0)}$ and $P_{\text{ur}}$ in Fig. \ref{fig6}. In Fig. \ref{fig7}, a comparison between the $T/T_{c,\text{nr}}$ dependence of $P_{\text{nr}}/n_{\text{tot}}T_{c,\text{nr}}$ of a rotating Bose gas in NR limit (dashed curve) with the  $T/T_{c,\text{ur}}^{(0)}$ dependence of $P_{\text{ur}}^{(0)}/n_{\text{tot}}T_{c,\text{ur}}^{(0)}$ of a nonrotating Bose gas in UR limit is made. The result shows that  a nonrelativistic Bose gas under rigid rotation behaves almost as a nonrotating gas in the UR limit. As it is shown in Fig. \ref{fig6}(a), at temperatures around the BEC critical temperature, the pressure of a rotating nonrelativistic gas increases comparing to a nonrotating nonrelativistic gas. The reason is the additional centrifugal force due to rotation in the rigidly rotating gas \cite{landau-book}.
%-----------------------------------------
\subsection{The heat capacity at $\boldsymbol{T\leq T_{c}}$ and $\boldsymbol{T>T_{c}}$}\label{sec5c}
%-----------------------------------------
To see how the temperature dependence of the fugacity at temperatures lower and higher than the critical temperature affects the behavior of various thermodynamic quantities, we compute in what follows the dependence of the heat capacity of a nonrotating and rotating Bose gas in NR and UR limits.
%-----------------------------------------
\subsubsection{Nonrotating Bose gas in NR and UR limits}\label{sec5c1}
%-----------------------------------------
Let us first consider a nonrotating Bose gas. The heat capacity is defined in \eqref{A29}, where $\epsilon$ is the energy density of the medium. Its thermal part is related to the thermal pressure through the corresponding EoS. In the case of a nonrelativistic and nonrotating bosonic medium, the EoS is given by \eqref{B18}. We thus have
\begin{eqnarray}\label{D33}
C_{V,\text{nr}}^{(0)}=\left(\frac{\partial\epsilon^{ (0)}_{\text{nr}}}{\partial T}\right)_{n}=\frac{3}{2}\left(\frac{\partial P^{ (0)}_{\text{nr}}}{\partial T}\right)_{n}.
\end{eqnarray}
Using the results for $P^{ (0)}_{\text{nr}}$ presented in \eqref{D29}, and having in mind that the $T$ dependence of the fugacity $z$ in this case is given by \eqref{D1} and \eqref{D8}, the heat capacity of a nonrotating Bose gas in NR limit reads
\begin{eqnarray}\label{D34}
C_{V,\text{nr}}^{(0)}&=&n_{\text{tot}}~\frac{15}{4}\frac{\zeta(5/2)}{\zeta(3/2)}\left(\frac{T}{T_{c,\text{nr}}^{(0)}}\right)^{3/2},\qquad \text{at $T<T_{c,\text{nr}}^{(0)}$},\nonumber\\
C_{V,\text{nr}}^{ (0)}&=&n_{\text{tot}}~\frac{15}{4}\frac{\zeta(5/2)}{\zeta(3/2)},\hspace{2.5cm}\text{at $T=T_{c,\text{nr}}^{(0)}$},\nonumber\\
C_{V,\text{nr}}^{ (0)}&=&n_{\text{tot}}~\left(\frac{15}{4}\frac{g_{5/2}(z)}{g_{3/2}(z)}-\frac{9}{4}\frac{g_{3/2}(z)}{g_{1/2}(z)}\right),\hspace{0.2cm} \text{at $T>T_{c,\text{nr}}^{(0)}$}.\nonumber\\
\end{eqnarray}
To derive the above results, \eqref{C8} and \eqref{D10} are also used. Let us notice that since $g_{1/2}(z=1)=\zeta(1/2)\to \infty$, the second term on the rhs of $C_{V,\text{nr}}^{ (0)}$ vanishes at $z=1$, or equivalently at $T=T_{c,\text{nr}}^{(0)}$. This is why $C_{V,\text{nr}}^{ (0)}$ is continuous at this temperature. It is also possible to determine $\partial C_{V,\text{nr}}^{(0)}/\partial T$ and show that it is discontinuous at $T=T_{c,\text{nr}}^{(0)}$ \cite{pathria-book},
\begin{eqnarray}\label{D35}
\left(\frac{\partial C_{V,nr}^{(0)}}{\partial T}\right)_{-}-
\left(\frac{\partial C_{V,nr}^{(0)}}{\partial T}\right)_{+}=\frac{27(\zeta(3/2))^{2}}{16\pi}\frac{n_{\text{tot}}}{T_{c,\text{nr}}^{(0)}},\nonumber\\
\end{eqnarray}
where subscripts $-$ and $+$ denote $T\to (T_{c,\text{nr}}^{(0)})_{-}$ and $T\to (T_{c,\text{nr}}^{(0)})_{+}$, respectively.
\par
For nonrotating Bose gas in UR limit, we have first to rewrite \eqref{D33} according to the EoS of this gas from \eqref{B31},
\begin{eqnarray}\label{D36}
C_{V,\text{ur}}^{(0)}=\left(\frac{\partial\epsilon^{ (0)}_{\text{ur}}}{\partial T}\right)_{n}=3\left(\frac{\partial P^{ (0)}_{\text{ur}}}{\partial T}\right)_{n}.
\end{eqnarray}
Plugging then the results from \eqref{D30} into \eqref{D36} and using the $T$ dependence of $\mathfrak{z}$ from \eqref{D1} and \eqref{D15} as well as \eqref{D16}, we obtain
\begin{eqnarray}\label{D37}
C_{V,\text{ur}}^{ (0)}&=&n_{\text{tot}}~12\frac{\zeta(4)}{\zeta(3)}\left(\frac{T}{T_{c,\text{ur}}^{(0)}}\right)^{3},\qquad \text{at $T<T_{c,\text{nr}}^{(0)}$},\nonumber\\
C_{V,\text{ur}}^{ (0)}&=&n_{\text{tot}}~12\frac{\zeta(4)}{\zeta(3)},\hspace{2.2cm}\text{at $T=T_{c,\text{ur}}^{(0)}$},\nonumber\\
C_{V,\text{ur}}^{ (0)}&=&n_{\text{tot}}~\left(12\frac{g_{4}(\mathfrak{z})}{g_{3}(\mathfrak{z})}-9\frac{g_{3}(\mathfrak{z})}{g_{2}(\mathfrak{z})}\right),\hspace{0.2cm} \text{at $T>T_{c,\text{ur}}^{(0)}$}.\nonumber\\
\end{eqnarray}
Unlike the result for a nonrotating Bose gas in NR limit, $C_{V,\text{ur}}^{ (0)}$ is discontinuous at $T=T_{c,\text{ur}}^{(0)}$ (or equivalently, at $\mathfrak{z}=1$).
%-----------------------------------------
\subsubsection{Rotating Bose gas in NR and UR limits}\label{sec5c2}
%-----------------------------------------
As it is shown in Sec. \ref{sec3a} and \ref{sec3b}, the EoS of rotating Bose gas in NR and UR limits are given by \eqref{C15} and \eqref{C29}. The heat capacities in these two cases thus read
\begin{eqnarray}\label{D38}
C_{V,\text{nr}}&=&\left(\frac{\partial\epsilon_{\text{nr}}}{\partial T}\right)_{n,\Omega}=\frac{5}{2}\left(\frac{\partial P_{\text{nr}}}{\partial T}\right)_{n,\Omega},\nonumber\\
C_{V,\text{ur}}&=&\left(\frac{\partial\epsilon_{\text{ur}}}{\partial T}\right)_{n,\Omega}=4\left(\frac{\partial P_{\text{ur}}}{\partial T}\right)_{n,\Omega},
\end{eqnarray}
where $P_{\text{nr}}$ and $P_{\text{ur}}$ at temperatures below and above the corresponding critical temperatures are given in \eqref{D31} and \eqref{D32}, respectively. Using \eqref{D38},
the heat capacity of a rotating Bose gas in NR limit reads
\begin{eqnarray}\label{D39}
C_{V,\text{nr}}&=&n_{\text{tot}}~\frac{35}{4}\frac{\zeta(7/2)}{\zeta(5/2)}\left(\frac{T}{T_{c,\text{nr}}}\right)^{5/2},\qquad~~\text{at $T<T_{c,\text{nr}}$},\nonumber\\
C_{V,\text{nr}}&=&n_{\text{tot}}~\frac{35}{4}\frac{\zeta(7/2)}{\zeta(5/2)},\hspace{2.5cm}\text{~~at $T=T_{c,\text{nr}}$},\nonumber\\
C_{V,\text{nr}}&=&n_{\text{tot}}~\left(\frac{35}{4}\frac{g_{7/2}(z)}{g_{5/2}(z)}-\frac{25}{4}\frac{g_{5/2}(z)}{g_{3/2}(z)}\right),\hspace{0.2cm} \text{at $T>T_{c,\text{nr}}$}, \nonumber\\
\end{eqnarray}
where at $T>T_{c,\text{nr}}$ the fugacity $z$ is given in \eqref{D20}. To derive $C_{V,\text{ur}}$, we also use $\left(\partial z/\partial T\right)_{n,\Omega}$ from \eqref{D22}.  In UR limit, the heat capacity is derived also from \eqref{D38}. At temperatures below and above the critical temperature, it is given by
\begin{eqnarray}\label{D40}
C_{V,\text{ur}}&=&n_{\text{tot}}~20\frac{\zeta(5)}{\zeta(4)}\left(\frac{T}{T_{c,\text{ur}}}\right)^{4},\qquad \text{at $T<T_{c,\text{nr}}^{(0)}$},\nonumber\\
C_{V,\text{ur}}&=&n_{\text{tot}}~20\frac{\zeta(5)}{\zeta(4)},\hspace{2.2cm}\text{at $T=T_{c,\text{ur}}$},\nonumber\\
C_{V,\text{ur}}&=&n_{\text{tot}}~\left(20\frac{g_{5}(\mathfrak{z})}{g_{4}(\mathfrak{z})}-16\frac{g_{4}(\mathfrak{z})}{g_{3}(\mathfrak{z})}\right),\hspace{0.2cm} \text{at $T>T_{c,\text{ur}}$}.\nonumber\\
\end{eqnarray}
At $T>T_{c,\text{ur}}$, the fugacity is given by \eqref{D27}. To derive \eqref{D40}, $\left(\partial{\mathfrak{z}}/\partial T\right)_{n,\Omega}$ from \eqref{D28} is used.
\par
In Fig. \ref{fig8}, the $T/T_{c}$ dependence of dimensionless heat capacity $C_{V}/n_{\text{tot}}$ is plotted for nonrelativistic [Fig. \ref{fig8}(a)] and ultrarelativistic [Fig. \ref{fig8}(b)] Bose gas with $\Omega=0$ (blue dashed curves) and $\Omega\neq 0$ (red solid curves). Except $C_{V,\text{nr}}^{(0)}$ which is continuous at the critical temperature, all the other curves exhibit certain discontinuity at the corresponding $T_{c}$s. According to Ehrenfest classification, these discontinuities may indicate a second order phase transition \cite{cond2006}. As it is mentioned in Sec. \ref{sec1}, similar results are previously presented in \cite{cond2006}, where the rotation is introduced using a completely different method.
\par
Let us notice that the horizontal dotted lines indicate the asymptotic values of $C_{V}$, according to Dulong-Petit law. They arises by taking the limit  $z\to 0$ in the expressions for the heat capacities at temperatures larger than the critical temperature in \eqref{D34}, \eqref{D37}, \eqref{D39}, and \eqref{D40}.
Using $g_{\nu}(z)\simeq z$ at $z\to 0$, $C_{V,\text{nr}}^{(0)}/n_{\text{tot}}\simeq 3/2 $, $C_{V,\text{ur}}^{(0)}/n_{\text{tot}}\simeq 3 $, $C_{V,\text{nr}}/n_{\text{tot}}\simeq 5/2 $, and $C_{V,\text{ur}}/n_{\text{tot}}\simeq 4 $. Notably, these values correspond to $c_{s}^{-2}$ appearing in the EoS $\epsilon=c_{s}^{-2}P$ of Bose gases in NR and UR limits with $\Omega=0$ and $\Omega\neq 0$.
%-----------------------------------------
\subsection{The angular momentum density of a rotating Bose gas at $\boldsymbol{T\leq T_{c}}$ and $\boldsymbol{T>T_{c}}$ in NR and UR limits}\label{sec5d}
\begin{figure}[hbt]
\includegraphics[width=8cm, height=6cm]{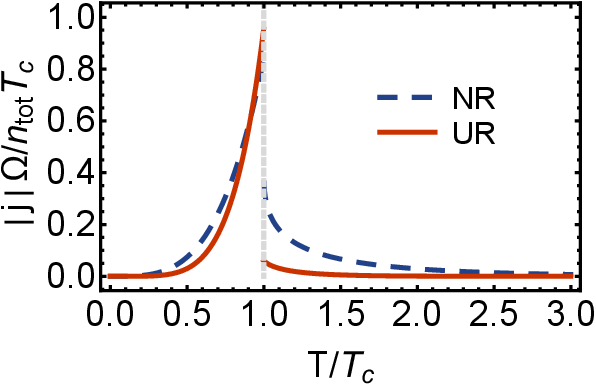}
\caption{color online. The $T/T_{c}$ dependence of the absolute value of dimensionless angular velocity density $|j|\Omega/n_{\text{tot}}T_{c}$ is plotted for nonrelativistic (blue dashed curves) and ultrarelativistic (red solid curves) Bose gas in the presence of rigid rotation [see \eqref{D41} and \eqref{D42}].  The vertical gray dashed line indicates the position of $T=T_{c}$ with $T_{c}\in\{T_{c,\text{nr}},T_{c,\text{ur}}\}$ for rotating Bose gas in NR and UR limits. In both cases, $|j|\Omega/n_{\text{tot}}T_{c}$ is discontinuous at $T=T_{c}$.}\label{fig9}
\end{figure}

%-----------------------------------------
Another important thermodynamic quantity associated with a rotating gas is the angular momentum density $j$, defined by \eqref{A26} in rotating Bose gases. Using the corresponding expression for the pressure in NR and UR limit at temperatures below, equal and above the critical temperature from \eqref{D31} and \eqref{D32}, we arrive at
\begin{eqnarray}\label{D41}
j_{\text{nr}}&=&-\frac{n_{\text{tot}}T_{c,\text{nr}}}{\Omega}~\frac{\zeta(7/2)}{\zeta(5/2)}\left(\frac{T}{T_{c,\text{nr}}}\right)^{7/2},\qquad \text{at $T<T_{c,\text{nr}}$},\nonumber\\
j_{\text{nr}}&=&-\frac{n_{\text{tot}}T_{c,\text{nr}}}{\Omega}~\frac{\zeta(7/2)}{\zeta(5/2)},\hspace{2.45cm}\text{at $T=T_{c,\text{nr}}$},\nonumber\\
j_{\text{nr}}&=&-\frac{n_{\text{tot}}T_{c,\text{nr}}}{\Omega}~\frac{g_{7/2}(z)}{g_{5/2}(z)}\frac{T}{T_{c,\text{nr}}}\left(1-\frac{(g_{5/2}(z))^{2}}{g_{3/2}(z)g_{7/2}(z)}\right),\nonumber\\
&&\hspace{5.4cm} \text{at $T>T_{c,\text{nr}}$},\nonumber\\
\end{eqnarray}
for the NR limit and
\begin{eqnarray}\label{D42}
j_{\text{ur}}^{ }&=&-\frac{n_{\text{tot}}T_{c,\text{ur}}}{\Omega}~\frac{\zeta(5)}{\zeta(4)}\left(\frac{T}{T_{c,\text{ur}}}\right)^{5},\qquad \text{at $T<T_{c,\text{ur}}$},\nonumber\\
j_{\text{ur}}^{ }&=&-\frac{n_{\text{tot}}T_{c,\text{ur}}}{\Omega}~\frac{\zeta(5)}{\zeta(4)},\hspace{2.22cm}\text{at $T=T_{c,\text{ur}}$},\nonumber\\
j_{\text{ur}}^{ }&=&-\frac{n_{\text{tot}}T_{c,\text{ur}}}{\Omega}~\frac{g_{5}(\mathfrak{z})}{g_{4}(\mathfrak{z})}\frac{T}{T_{c,\text{ur}}}\left(1-\frac{(g_{4}(\mathfrak{z}))^{2}}{g_{3}(\mathfrak{z})g_{5}(\mathfrak{z})}\right),\nonumber\\
&&\hspace{4.8cm} \text{at $T>T_{c,\text{ur}}$},\nonumber\\
\end{eqnarray}
for the UR limit. Note that the fugacities $z$ and $\mathfrak{z}$, appearing in \eqref{D41} and \eqref{D42} at $T>T_{c,\text{nr}}$ are given in \eqref{D20} and \eqref{D27}. Moreover, $\left(\partial{z}/\partial\Omega\right)_{n,T}$ and $\left(\partial\mathfrak{z}/\partial\Omega\right)_{n,T}$ from \eqref{D22} and \eqref{D28} are also used. According to above results, in NR as well as UR limits, the angular momenta at $T=T_{c}$ and $T>T_{c}$ are discontinuous.
In Fig. \ref{fig9}, the $T/T_{c}$ dependence of dimensionless absolute value of the angular momentum density is plotted for nonrelativistic and ultrarelativistic Bose gases under rotation. The results arise from \eqref{D41} and \eqref{D42}. Because of the behavior of $P$ at temperatures around $T_{c}$, the angular momentum density exhibits a certain discontinuity at $T=T_{c}$ indicated with a vertical gray line in Fig. \ref{fig9}. It is noteworthy that at $z\to 0$ (high temperatures) $j_{\text{nr}}$ as well as $j_{\text{ur}}$ vanish [see \eqref{D41} and \eqref{D42} and use $g_{\nu}(z)\simeq z$ at $z\to 0$].
%-----------------------------------------
\subsection{The entropy density and latent heat at $\boldsymbol{T=T_{c}}$}\label{sec5e}
%-----------------------------------------
In this part, we first use \eqref{A28} to compute the entropy density $s$ for nonrotating and rotating Bose gases in NR and UR limits. In particular, the corresponding expressions for $\epsilon, P, n$, and $j$ at critical temperatures in these limits are utilized.\footnote{For nonrotating Bose gas, we set $j=0$ in \eqref{A28}.} Then, the latent heat of these gases at critical temperatures corresponding to NR and UR limits are determined from \cite{rebhan-book}
\begin{eqnarray}\label{D43}
q\equiv T_{c}\frac{s(T=T_{c})}{n(T=T_{c})}.
\end{eqnarray}
\begin{figure}[hbt]
\includegraphics[width=8cm, height=6cm]{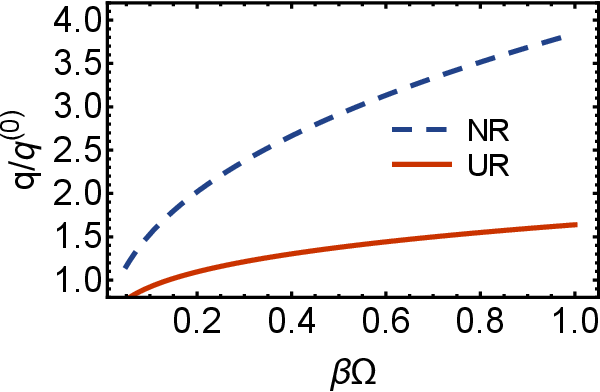}
\caption{color online. The $\beta\Omega$ dependence of the ratio $q/q^{(0)}$ for a nonrelativistic (dashed curve) and ultrarelativistic (solid curve) is plotted. For all values of $\beta\Omega<1$, the latent heat of a rotating Bose gas  in NR limit is larger than the one in UR limit. Moreover, independent of $\beta\Omega$, the latent heat of a gas under rotation is, in general, larger than that of a nonrotating gas. }\label{fig10}
\end{figure}
%-----------------------------------------
\subsubsection{Nonrotating Bose gas in NR and UR limits}\label{sec6d1}
%-----------------------------------------
To determine the entropy density of a nonrotating gas in NR limit, $s_{\text{nr}}$, let us consider $P_{\text{nr}}^{(0)}$ at $T=T_{c,\text{nr}}^{(0)}$ from \eqref{D29}. Using the EoS \eqref{B18}, and the fact that $n_{\text{nr}}^{(0)}=n_{\text{tot}}$ at $T=T_{c,\text{nr}}^{(0)}$ [see \eqref{C9}], we arrive at
\begin{eqnarray}\label{D44}
q_{\text{nr}}^{(0)}=\frac{5}{2}\frac{\zeta(5/2)}{\zeta(3/2)}T_{c,\text{nr}}^{(0)}-m,
\end{eqnarray}
upon using $s_{\text{nr}}^{(0)}=(\epsilon_{\text{nr}}^{(0)}+P_{\text{nr}}^{(0)}-m n_{\text{tot}})/T_{c,\text{nr}}^{(0)}$. Note that at $T=T_{c,\text{nr}}^{(0)}$, we have $\mu=m$. Neglecting the rest mass $m$, the latent heat for nonrelativistic and nonrotating Bose gas turns out to be positive.
\par
Similarly, the latent heat of a nonrotating Bose gas in UR limit is given in terms of $P_{\text{ur}}^{(0)}$ and $T_{c,\text{ur}}^{(0)}$ from \eqref{B29} and \eqref{C11} as well as the EoS \eqref{B31}. Plugging these expressions into $s_{\text{ur}}^{(0)}=(\epsilon_{\text{ur}}^{(0)}+P_{\text{ur}}^{(0)}-m n_{\text{tot}})/T_{c,\text{ur}}^{(0)}$, we obtain
\begin{eqnarray}\label{D45}
q_{\text{ur}}^{(0)}=4\frac{\zeta(4)}{\zeta(3)}T_{c,\text{ur}}^{(0)},
\end{eqnarray}
which is always positive.
%-----------------------------------------
\subsubsection{Rotating Bose gas in NR and UR limits}\label{sec6d2}
%----------------------------------------
The entropy density of a rotating Bose gas in NR limit at $T_{c,\text{nr}}$ can be derived from
$s_{\text{nr}}=(\epsilon_{\text{nr}}+P_{\text{nr}}-m n_{\text{tot}})/T_{c,\text{nr}}$, where $\epsilon_{\text{nr}}=5/2P_{\text{nr}}$ [see the EoS from \eqref{B15}], with  $P_{\text{nr}}$ and from \eqref{D31}. The latent heat of a nonrelativistic Bose gas under rigid rotation is thus given by
\begin{eqnarray}\label{D46}
q_{\text{nr}}=\frac{7}{2}\frac{\zeta(7/2)}{\zeta(5/2)}T_{c,\text{nr}}-m.
\end{eqnarray}
Assuming $m\sim 0$ in \eqref{D44} and \eqref{D46}, the ratio $q_{\text{nr}}/q_{\text{nr}}^{(0)}$ reads
\begin{eqnarray}\label{D47}
\frac{q_{\text{nr}}}{q_{\text{nr}}^{(0)}}=\frac{7}{5}\frac{\zeta(7/2)\zeta(3/2)}{(\zeta(5/2))^{2}}\frac{T_{c,\text{nr}}}{T_{c,\text{nr}}^{(0)}}\sim 2.3 \frac{T_{c,\text{nr}}}{T_{c,\text{nr}}^{(0)}},
\end{eqnarray}
where the relation between  $T_{c,\text{nr}}$ and $T_{c,\text{nr}}^{(0)}$ is given in \eqref{C23} and depends explicitly on the angular velocity $\Omega$.
\par
As concerns the latent heat of a rotating Bose gas in UR limit, it is defined in terms of entropy $s_{\text{ur}}=(\epsilon_{\text{ur}}+P_{\text{ur}})/T_{c,\text{ur}}$. Plugging $P_{\text{ur}}$ at $T=T_{c,\text{ur}}$ from \eqref{D32} into this expression, we arrive at
\begin{eqnarray}\label{D48}
q_{\text{ur}}=5\frac{\zeta(5)}{\zeta(4)}T_{c,\text{ur}}.
\end{eqnarray}
Comparing \eqref{D45} and \eqref{D48}, we thus obtain
\begin{eqnarray}\label{D49}
\frac{q_{\text{ur}}}{q_{\text{ur}}^{(0)}}=\frac{5}{4}\frac{\zeta(5)\zeta(3)}{(\zeta(4))^{2}}\frac{T_{c,\text{ur}}}{T_{c,\text{ur}}^{(0)}}\sim 1.3 \frac{T_{c,\text{ur}}}{T_{c,\text{ur}}^{(0)}}.
\end{eqnarray}
Here, the relation between  $T_{c,\text{ur}}$ and $T_{c,\text{ur}}^{(0)}$ is given in \eqref{C32} and depends explicitly on the angular velocity $\Omega$. The $\beta\Omega$ dependence of the ratio $q_{\text{nr}}/q_{\text{nr}}^{(0)}$ and   $q_{\text{ur}}/q_{\text{ur}}^{(0)}$ from \eqref{D47} and \eqref{D49} is plotted in Fig. \ref{fig10}. According to these results,
for all values of $\beta\Omega<1$, the latent heat of a gas under rotation is, in general, larger than that of a nonrotating gas. Moreover, independent of $\beta\Omega$, the ratio $q_{\text{ur}}/q_{\text{ur}}^{(0)}$ is in general smaller than $q_{\text{nr}}/q_{\text{nr}}^{(0)}$.
%----------------------------------------
\section{Concluding remarks}\label{sec6}
%----------------------------------------
In the present work, we studied the impact of rigid rotation on the BE transition in nonrelativistic and ultrarelativistic Bose gases. To this purpose, we introduced the rotation in the Lagrangian density of a free complex scalar field and computed the grand canonical partition function of this model at finite temperature and density. By assuming slow rotation, we were able to perform the summation over the quantum number $\ell$ that arises due to the rotational motion and the formulation of the problem in a cylindrical coordinate system. This method enabled us to derive analytical results for the pressure, $P$, energy and number densities, $\epsilon$ and $n$ and  compare them with the corresponding results for nonrotating Bose gases in NR and UR limits.
\par
We also focused on the BEC phenomenon in rotating Bose gas. We investigated, in particular, how the critical temperature of the BE transition, as well as the condensate and thermal fractions, are affected by rigid rotation. We derived analytical expressions for these quantities and compared them with the corresponding expressions for a nonrotating Bose gas in both NR and UR limits. Our findings revealed that the critical exponents associated with the condensate and thermal fractions differ from those in a nonrotating Bose gas. Interestingly, all these exponents are related to $c_{s}^{-2}$ of the corresponding models, where $c_{s}$ is the speed of sound in the EoS of the Bose gas, $\epsilon=c_{s}^{-2} P$.  We showed that rigid rotation with small angular velocity decreases the speed of sound and the critical temperature of the BE transition of a nonrotating Bose gas. These results suggest that a rotating nonrelativistic Bose gas behaves similarly as a nonrotating ultrarelativistic one.
\par
We finally determined various thermodynamic quantities such as pressure, heat capacity, and angular momentum density at temperatures both below and above the critical temperature of the BE transition $T_{c}$.
The key component in this calculation was the fugacities corresponding to nonrelativistic and ultrarelativistic Bose gases under rigid rotation, as all of the above quantities explicitly depend on it. Unlike the fugacity in a nonrotating Bose gas within these limits, the fugacity in rotating gases depends not only on temperature but also on the angular velocity. We determined its temperature dependence both below and above $T_{c}$ for vanishing and nonvanishing angular velocities, and showed how the angular velocity affects this dependence. The comparison made in Fig. \ref{fig7} between the pressures of a rotating gas in NR limit and a nonrotating gas in UR limit, underline the statement that a nonrelativistic Bose gas under rigid rotating behaves similar to a nonrotating gas in ultrarelativistic limit, as indicated above. To explain this effect, it is possible to compare the energy dispersion relation $\epsilon_{p,\text{ur}}\simeq -\Omega|\ell|+p$ and $\epsilon_{p,\text{nr}}\simeq -\Omega|\ell|+m+p^{2}/2m$ of a rotating Bose gas in UR and NR limits with the energy dispersion relation of phonons and rotons in a superfluid medium, $\epsilon_{p,\text{phonon}}=c_{s}p$ and $\epsilon_{p,\text{roton}}=\Delta+(p-p_{0})^{2}/2m$. In the absence of interaction, $\Omega|\ell|$ plays the role of the mass gap $\Delta$ in $\epsilon_{p,\text{roton}}$.\footnote{We remind that $m$ in $\epsilon_{p,\text{nr}}$ is already considered in the fugacity $z$ from \eqref{B1}.}  Once $p_{0}=0$ and the mass gap $\Omega|\ell|$ is small enough (slow rotation), $\epsilon_{p,\text{ur}}(\Omega=0)$ and $\epsilon_{p,\text{nr}}$ almost coincide at $p\approx  0$.
\par
Once we compare the $T$ dependence of the heat capacities ($C_{V}$s) corresponding to rotating and nonrotating gases, we observe that whereas the $C_{V}$ of a nonrotating Bose gas in NR limit is continuous and possesses a nonanalyticity at the corresponding critical temperature, the heat capacity of a rotating gas in the same limit is discontinuous at this temperature, similar to the heat capacity of a nonrotating gas in UR limit. This indicates that the rotation may change the feature of the BEC phase transition of a nonrelativistic Bose gas. It is noteworthy that similar results are previously obtained in \cite{cond2006}, in which rotation is introduced in a completely different manner as the one presented in the present paper. The same discontinuity as the one appearing in the heat capacity at $T_{c}$ occurs in the angular velocity density of rotating Bose gas at the critical temperature in both NR and UR limits.
We also compute the latent heat which is necessary to convert the condensate to thermal phase at the critical temperature and show that at this temperature the latent heat for a rotating Bose gas is, in general, larger than the latent heat of a nonrotating gas.
\par
As previously mentioned, the results presented in this work have applications across various branches of theoretical physics, ranging from condensed matter to astroparticle physics. The computation carried out in this paper can be extended by introducing a constant background magnetic field and to investigate the combined effects of the magnetic field and the rigid rotation on the BEC phenomenon. A preliminary attempt in this direction is made recently in \cite{Voskresensky:2024vfx}, which explores the possibility of charged pion condensation in the presence of rotation and external electromagnetic fields. Apart from this work, the phase structure of a relativistic self-interacting boson gas has been studied recently in \cite{karpenko2024}. It is shown that BEC significantly increases the critical temperature of the liquid-gas phase transition. Specifically, the sensitivity of the critical point of the mixed phase to the presence of BEC is explored in this work. The results have implications for the physics of neutron stars as well as for heavy ion collisions at high densities. Here, rotation plays a crucial role. It would be intriguing to further investigate the effect of rotation on this phase diagram.
\begin{appendix}
\section{Useful formula}\label{appA}
\setcounter{equation}{0}
%%%%%%%%%%%%%%%%%%%%%%%%%
To arrive at \eqref{A19}, we used the orthonormality relations
\begin{eqnarray}\label{appA1}
\int_{0}^{\beta} d\tau e^{i(\omega_{n}-\omega_{n'})\tau} &=&\beta\delta_{n,n'},\nonumber\\
\int_{0}^{2\pi} d\varphi e^{i(\ell-\ell')\varphi} &=&2\pi\delta_{\ell,\ell'},\nonumber \\
\int_{-\infty}^{\infty} dz e^{i(k_{z}-k'_{z})z} &=& 2\pi\delta(k_{z}-k'_{z}),\nonumber\\			\int_{0}^{\infty} dr\, r J_{\ell}(k_{\perp}r)  J_{\ell}(k'_{\perp}r) &=&\frac{1}{k_{\perp}} \delta(k_{\perp}-k'_{\perp}),
\end{eqnarray}
and obtained
\begin{eqnarray}\label{appA2}
\lefteqn{\hspace{-1.5cm}\int_{X} e^{i(\omega_{n}-\omega_{n'})\tau}  e^{i(\ell-\ell')\varphi} e^{i(k_{z}-k'_{z})z} J_{\ell}(k_{\perp}r)  J_{\ell}(k'_{\perp}r)}\nonumber\\
&=& \beta (2\pi)^{2} \hat{\delta}_{\ell,\ell'}^{n,n'}(k_{z},k_{\perp};k'_{z},k'_{\perp}),
\end{eqnarray}
with $\hat{\delta}_{\ell,\ell'}^{n,n'}$ defined as
\begin{eqnarray}\label{appA3}
\hat{\delta}_{\ell,\ell'}^{n,n'}(k_{z},k_{\perp};k'_{z},k'_{\perp})\equiv \frac{1}{k_{\perp}} \delta(k_{z}-k'_{z}) \delta(k_{\perp}-k'_{\perp}) \delta_{n,n'} \delta_{\ell,\ell'}.\nonumber\\
\end{eqnarray}
Moreover, we used the recursion relation
	\begin{eqnarray}\label{appA4}
		J_{\ell+1}(k_{\perp}r) = \frac{2\ell}{k_{\perp}r}J_{\ell}(k_{\perp}r)-J_{\ell-1}(k_{\perp}r),
	\end{eqnarray}
and the derivative
	\begin{eqnarray}\label{appA5}
		\frac{\partial J_{\ell}(k_{\perp}r)}{\partial r}
		= \frac{k_{\perp}}{2} \left( J_{\ell-1}(k_{\perp}r) - J_{\ell+1}(k_{\perp}r)  \right),
	\end{eqnarray}
for the Bessel function $J_{\ell}(k_{\perp}r)$.
%-----------------------------------------
\section{Useful integrals}\label{appB}
\setcounter{equation}{0}
%-----------------------------------------
In this appendix, we perform the integration
\begin{eqnarray}\label{appB1}
I=\int d\tilde{k}~e^{-\beta j\omega_{k}}, \quad\text{with}\quad d\tilde{k}=\frac{dk_{z}k_{\perp}dk_{\perp}}{(2\pi)^{2}},
\end{eqnarray}
for two different NR and UR limits.
%-----------------------------------------
\subsection{Nonrelativistic limit}\label{appB1}
%-----------------------------------------
Let us consider
\begin{eqnarray}\label{appB2}
I_{\text{nr}}=\int d\tilde{k}~e^{-\beta j \omega_{k} },
\end{eqnarray}
with $\omega_{k}\approx\frac{k^{2}}{2m}$ and $k=(k_{\perp}^{2}+k_{z}^{2})^{1/2}$ in a cylindrical coordinate system. Using
\begin{eqnarray}\label{appB3}
\int_{-\infty}^{\infty}dk_{z}~e^{-\alpha k_{z}^2}=\left(\frac{\pi}{\alpha}\right)^{1/2}~~\mbox{and}~~
\int_{0}^{\infty}k_{\perp}dk_{\perp}~e^{-\alpha k_{\perp}^{2}}=\frac{1}{2\alpha},\nonumber\\
\end{eqnarray}
for $\mbox{Re}(\alpha)>0$, the Gaussian integration over $k_{z}$ and $k_{\perp}$ results in
\begin{eqnarray}\label{appB4}
\int\frac{dk_{z}k_{\perp}dk_{\perp}}{(2\pi)^{2}}e^{-\alpha(k_{z}^{2}+k_{\perp}^{2})}=\left(\frac{1}{4\pi\alpha}\right)^{3/2}.
\end{eqnarray}
For $\alpha=\frac{j}{2mT}$, we obtain
\begin{eqnarray}\label{appB5}
I_{\text{nr}}=\frac{1}{\lambda_{T}^{3} j^{3/2}},
\end{eqnarray}
with the thermal length $\lambda_{T}$ defined in \eqref{B9}.
%-----------------------------------------
\subsection{Ultrarelativistic limit}\label{appB2}
%-----------------------------------------
Let us consider
\begin{eqnarray}\label{appB6}
I_{\text{ur}}=\int d\tilde{k}~e^{-\beta j \omega_{k}},
\end{eqnarray}
with $\omega_{k}\approx k$. Following the method presented in \cite{sadooghi2024}, and using the Mellin transformation of $e^{-\beta \omega_{k} j}$, we arrive first at
\begin{eqnarray}\label{appB7}
\hspace{-0.5cm}e^{-\beta \omega_{k} j}=\frac{1}{2\pi i}\int_{c-i\infty}^{c+i\infty}dz\Gamma(z)(\beta j)^{-z}(\omega_{k}^{2})^{-z/2}.
\end{eqnarray}
Plugging then
\begin{eqnarray}\label{appB8}
(\omega_{k}^{2})^{-z/2}=\frac{1}{\Gamma(z/2)}\int_{0}^{\infty}dt~t^{z/2-1}e^{-\omega_{k}^{2} t},
\end{eqnarray}
into \eqref{appB7}, we obtain
\begin{eqnarray}\label{appB9}
\lefteqn{I_{\text{ur}}=\lim\limits_{\Lambda\to 0}\frac{1}{2\pi i}\int_{c-\infty}^{c+\infty}dz\frac{\Gamma(z)}{\Gamma(z/2)} (\beta j)^{-z}
}\nonumber\\
&&\times \int_{0}^{\infty}dt~ t^{z/2-1}e^{-\Lambda^{2}t}\int d\tilde{k}~e^{-(k_{z}^{2}+k_{\perp}^{2}) t}.
\end{eqnarray}
Here, we have introduced a regulator $\Lambda$. We continue by performing the integrations over $\boldsymbol{k}, t$, and $z$, and eventually take the limit $\Lambda\to 0$. We use \eqref{appB4} to evaluate the Gaussian integration over $k_{z}$ and $k_{\perp}$. Plugging \eqref{appB4} into \eqref{appB9}, using
\begin{eqnarray}\label{appB10}
\int_{0}^{\infty}dt~ t^{-5/2+z/2} e^{-\Lambda^{2}t}=\Lambda^{3-z}\Gamma\left(\frac{z-3}{2}\right),
\end{eqnarray}
and the Legendre formula for $\Gamma(z)$,
\begin{eqnarray}\label{appB11}
\Gamma(z)=\frac{2^{z}}{2\pi^{1/2}}\Gamma\left(\frac{z}{2}\right)\Gamma\left(\frac{z+1}{2}\right),
\end{eqnarray}
we arrive at
\begin{eqnarray}\label{appB12}
I_{\text{ur}}&=&\lim\limits_{\Lambda\to 0}\frac{\Lambda^{3}}{16\pi^{2}}\nonumber\\
&&\times\frac{1}{2\pi i}\int_{c-i\infty}^{c+i\infty}dz\Gamma\left(\frac{z-3}{2}\right)\Gamma\left(\frac{z+1}{2}\right)\left(\frac{\Lambda\beta}{2}\right)^{-z}. \nonumber\\
\end{eqnarray}
Performing the Mellin-Barnes type integral over $z$ by making use of
\begin{eqnarray}\label{appB13}
\frac{1}{2\pi i}\int_{c-i\infty}^{c+i\infty}dz\Gamma\left(\frac{z-3}{2}\right)\Gamma\left(\frac{z+1}{2}\right)\left(\frac{y}{2}\right)^{-z}=\frac{8}{y}K_{2}(y),\nonumber\\
\end{eqnarray}
 and taking eventually the limit $y=\Lambda\beta\to 0$, we finally obtain
\begin{eqnarray}\label{appB14}
I_{\text{ur}}=\frac{1}{\Lambda_{T}^{3}j^{3}},
\end{eqnarray}
with $\Lambda_{T}$ defined in \eqref{B24}. Here,
\begin{eqnarray}\label{appB15}
K_{2}(x)\xrightarrow{x\to 0}\frac{2}{x},
\end{eqnarray}
is also used.
\end{appendix}
%%%% %%%%%%%%%%%%%%%%%%%%%

\end{document}